\documentclass[fleqn]{aa}

\makeatletter
\renewcommand*\aa@pageof{, page \thepage{} of \pageref*{LastPage}}
\makeatother

\bibpunct{(}{)}{;}{a}{}{,}

\usepackage[varg]{txfonts}  

\usepackage[T1]{fontenc}
\usepackage{verbatim}
\usepackage{multirow, booktabs}
\usepackage{fp}	%	To evaluate results in macros

\usepackage{lscape}

\usepackage{graphicx}	% Including figure files

\usepackage{amsmath}	% Advanced maths commands

\usepackage{amssymb}	% Extra maths symbols

\usepackage{xspace}

\usepackage[dvipsnames]{xcolor}

\usepackage[colorlinks,citecolor=blue,urlcolor=blue,linkcolor=blue,bookmarks=false,breaklinks,hypertexnames=true]{hyperref}

\newcommand{\chandra}{{\em Chandra}}

\newcommand{\nustar}{{\em NuSTAR}}

\newcommand{\xmmn}{{\em XMM-Newton}}

\newcommand{\ero}{\textit{eROSITA}}
\newcommand{\srg}{\textit{Spectrum Roentgen Gamma}}

\newcommand{\mstar}{$M_{\star}$}

\newcommand{\msun}{M$_{\odot}$}	%	M$_{\Sun}$

\newcommand{\lx}  {{$L_{\rm{X}}$}}

\newcommand{\nh} {{$N_{\rm{H}}$}}

\newcommand{\sfr}  {{M$_{\odot}$ yr$^{-1}$}}
\newcommand{\es}  {{erg s$^{-1}$}}
\newcommand{\esc}  {{erg s$^{-1}$ cm$^{-2}$}}

\newcommand{\ten}[2]{$#1\times 10^{#2}$}	%	{value}x10^{power}

\newcommand{\dtf}  {{D$_{25}$}}

\newcommand{\nefeds}{32\,684\xspace}
\newcommand{\nefedspe}{32\,646\xspace}
\newcommand{\hecgal}{1181}	% # of HECATE galaxies in eFEDS FOV

\newcommand{\hgalssfr}{409\xspace}	% # of HECATE galaxies in eFEDS FOV with sSFR estimates

\newcommand{\fov}{140\xspace}

\newcommand{\nmat}{94\xspace}
\newcommand{\sfgal}{36\xspace}	% late-type galaxies (40)

\newcommand{\normgal}{37\xspace} 
\newcommand{\egal}{1\xspace}	% early-type galaxies

\newcommand{\nsdss}{59}
\newcommand{\nsdssgal}{57}

\newcommand{\nspec}{76}

\newcommand{\metng}{27}	%	# of star-forming galaxies with metallicities

\newcommand{\eg}{e.g.\xspace}
\newcommand{\ie}{i.e.\xspace}

\graphicspath{{./}{figures/}}

\begin{document}

\title{The \ero\ Final Equatorial-Depth Survey (eFEDS):}
\subtitle{Presenting The Demographics of X-ray Emission From Normal Galaxies}

\author{N. Vulic\inst{\ref{nv1},\ref{nv2},\ref{nv3},\ref{nv4}} \and
A. E. Hornschemeier\inst{\ref{nv1},\ref{ah1}} \and
F. Haberl\inst{\ref{mpe}} \and
A. R. Basu-Zych\inst{\ref{nv1},\ref{arb2}} \and
E. Kyritsis\inst{\ref{az1},\ref{az3}} \and
A. Zezas\inst{\ref{az1},\ref{cfa},\ref{az3}} \and
M. Salvato\inst{\ref{mpe}} \and
A. Ptak\inst{\ref{nv1},\ref{ah1}} \and
A. Bogdan\inst{\ref{cfa}} \and
K. Kovlakas\inst{\ref{az1},\ref{az3}} \and
J. Wilms\inst{\ref{jw}} \and
M. Sasaki\inst{\ref{jw}} \and
T. Liu\inst{\ref{mpe}} \and
A. Merloni\inst{\ref{mpe}} \and
T. Dwelly\inst{\ref{south}} \and
H. Brunner\inst{\ref{mpe}} \and
G. Lamer\inst{\ref{aip}} \and
C. Maitra\inst{\ref{mpe}} \and 
K. Nandra\inst{\ref{mpe}} \and
A. Santangelo\inst{\ref{iaat}}
}

\institute{
NASA Goddard Space Flight Center, Code 662, Greenbelt, MD 20771, USA \\
\email{nvulic@uwo.ca}\label{nv1} \and
Department of Astronomy, University of Maryland, College Park, MD 20742-2421, USA \label{nv2} \and
Center for Research and Exploration in Space Science and Technology, NASA/GSFC, Greenbelt, MD 20771, USA \label{nv3} \and
Eureka Scientific, Inc., 2452 Delmer Street, Suite 100, Oakland, CA 94602-3017, USA \label{nv4} \and
Department of Physics and Astronomy, Johns Hopkins University, 3400 N. Charles Street, Baltimore, MD 21218, USA \label{ah1} \and
Max-Planck-Institut f{\"u}r extraterrestrische Physik, Gie{\ss}enbachstra{\ss}e 1, 85748 Garching, Germany \label{mpe} \and
Department of Physics, University of Maryland Baltimore County, Baltimore, MD 21250, USA \label{arb2} \and
Physics Department \& Institute of Theoretical \& Computational Physics, University of Crete, 71003 Heraklion, Crete, Greece \label{az1} \and
Harvard-Smithsonian Center for Astrophysics, 60 Garden Street, Cambridge, MA 02138, USA \label{cfa} \and
Institute of Astrophysics, Foundation for Research and Technology-Hellas, GR-71110 Heraklion, Greece \label{az3}\and
Dr.\ Karl Remeis-Observatory and Erlangen Centre for Astroparticle Physics, Friedrich-Alexander-Universit\"at Erlangen-N\"urnberg, Sternwartstr.~7, 96049 Bamberg, Germany \label{jw} \and
Department of Physics and Astronomy, University of Southampton, Southampton SO17 1BJ, UK \label{south} \and
Leibniz Institut f\"ur Astrophysik, An der Sternwarte 16, 14482 Potsdam, Germany \label{aip} \and
Institut f{\"u}r Astronomie und Astrophysik T{\"u}bingen, Universit{\"a}t T{\"u}bingen, Sand 1, 72076 T{\"u}bingen, Germany \label{iaat}
}

\titlerunning{eFEDS Normal Galaxies}
\authorrunning{N. Vulic}

\label{firstpage}

\date{Received date /
Accepted date }

\abstract{% Context
The \ero\ Final Equatorial Depth Survey (eFEDS), completed in survey mode during the calibration and performance verification phase of the \ero\ instrument on \srg, delivers data at and beyond the final depth of the four-year \ero\ all-sky survey (eRASS:8), $f_{0.5-2\,\text{ keV}}$ = \ten{1.1}{-14} \esc, over \fov deg$^{2}$. It provides the first view of normal galaxy X-ray emission from X-ray binaries (XRBs) and the hot interstellar medium at the full depth of eRASS:8. 
}
{%  Aims
We use the Heraklion Extragalactic Catalogue (HECATE) of galaxies to correlate with eFEDS X-ray sources and identify X-ray detected normal galaxies. This flux-limited X-ray survey is relatively free from selection effects and enables the study of integrated normal galaxy X-ray emission and its relation to galaxy parameters such as stellar mass, star formation rate, and metallicity. 
}
{%  Methods
We cross-correlate \nefedspe eFEDS X-ray sources to \hecgal\ HECATE normal galaxies and obtain \nmat matches. We classify galaxies as star-forming, early-type, composite, and AGN using SDSS and 6dF optical spectroscopy. 
}
{%  Results
The eFEDS field harbours \normgal normal galaxies: \sfgal late-type (star-forming) galaxies and \egal early-type galaxy. There are 1.9 times as many normal galaxies as predicted by scaling relations via SIXTE simulations, with an overabundance of late-type galaxies and a dearth of early-type galaxies. 
Dwarf galaxies with high specific star formation rate (SFR) have elevated \lx/SFR when compared with specific SFR and metallicity, indicating an increase in XRB emission due to low-metallicity. We expect that eRASS:8 will detect 12\,500 normal galaxies, the majority of which will be star-forming, with the caveat that there are unclassified sources in eFEDS and galaxy catalogue incompleteness issues that could increase the actual number of detected galaxies over these current estimates. 
}
{%  Conclusions
 eFEDS observations detected a rare population of galaxies -- the metal-poor dwarf starbursts -- that do not follow known scaling relations. eRASS is expected to discover significant numbers of these high-redshift analogues, which are important for studying the heating of the intergalactic medium at high-redshift. Further investigation of the hot gas emission from normal galaxies and stochastic effects in the dwarf galaxy population are required to constrain their X-ray output.
}


\maketitle

\section{Introduction} \label{sec:intro}

Understanding the nature of X-ray emission from normal galaxies, i.e., galaxies without an active galactic nucleus (AGN), is critically important for several reasons.  First, much of this X-ray emission arises from accretion onto compact objects, with the dominant accreting compact object population, in the absence of an accreting super-massive black hole (SMBH), consisting of neutron stars (NS) and stellar-mass black holes (BH). Accreting NS and BH uniquely trace the endpoints of massive star formation and are observable as X-ray Binaries (XRBs). In addition to including the progenitor population for gravitational waves detected by, e.g., Advanced LIGO \citep{abbott09-19},  the collective X-ray output from XRBs can rival that of accreting SMBH (AGN) at the critical epochs of reionization and Cosmic Dawn ($6 \lesssim z \lesssim 20$) when the  first galaxies in the Universe were forming \citep{fragos02-13,mesinger04-14,pacucci09-14}.  These lower-mass accreting systems effectively ``outshine'' their SMBH counterparts, playing a possibly significant role in heating the primordial intergalactic medium \citep{madau05-17}.  Given the difficulty of directly observing X-ray emission at high-redshifts, however, it is \emph{local} studies of X-ray emission from galaxies that give us the best chance to characterize this emission.

Much of the X-ray emission from massive normal galaxies arises from the astrophysically important hot phase of the interstellar medium (ISM) in galaxies. The thermal structure of the ISM gives key information about various physical processes affecting galaxy evolution \citep[e.g.,][]{kim10-15}. Hot gas properties are expected to differ based on, e.g., mergers, ram-pressure stripping, and both AGN and stellar feedback. Additionally, there is a long-standing prediction that galaxies should contain hot gaseous halos, the Circumgalactic Medium (CGM). While there have been some individual detections of bright instances of such halos \citep{bogdan08-13, li12-17} around spiral galaxies, detecting such X-ray emission from a large sample of spiral galaxies remains elusive.

All of our current understanding of X-ray emission from normal galaxies, however, is based on relatively small numbers of galaxies as compared to the powerhouse wide-field surveys in the optical/IR such as the Sloan Digital Sky Survey (SDSS). The treasure trove of galaxy information in SDSS, for example, is contained within a population with peak distance $z\sim0.07$ ($d\sim 315$\,Mpc). What is thus needed is an X-ray survey scanning thousands of square degrees with sensitivity to galaxies at those sorts of distances. The  \ero\ All Sky Survey (eRASS) provides this sensitivity over a wide area for detecting normal galaxies \citep{basu-zych08-20}. In eRASS, \ero, which is the 0.3--10\,keV X-ray instrument aboard the \srg\ observatory, will cover the entire sky to moderate depth and at moderate angular resolution\footnote{Pre-launch estimate from \citet{merloni09-12} was 28\arcsec.} \citep[$\sim 10^{-14}\,\mathrm{erg}\,\mathrm{cm}^{-2}\,\mathrm{s}^{-1}$ and 26\arcsec; see details in][]{predehl03-21,merloni09-12}. Due to the relative proximity of the galaxies detected in eRASS, the X-ray data may be combined with estimates for the star formation rate (SFR), stellar masses, and  metallicities in spatially-resolved galaxies that are better-characterized than high-$z$ populations.  Galaxy samples detected in the X-ray band will finally reach much larger numbers (thousands of galaxies versus 10s to 100s) with eRASS:8. Additionally, the \ero\ sample will be free from the selection effects of targeted observations that distort current galaxy samples. 
 
eRASS observations began in 2019 December and will continue through 2023 December, with the full sky scanned every six months for a total of eight iterations of eRASS to reach the final depth (eRASS:1 -- eRASS:8). However, during the Performance Verification (PV) phase of the mission in mid-to-late 2019, the \ero\ Final Equatorial-Depth Survey (eFEDS) was conducted over \fov deg$^{2}$ at and beyond the ultimate eRASS:8 depth to provide a pilot program for the eRASS survey.

In this paper, we investigate the properties of $\sim100$ galaxies that are part of the Heraklion Extragalactic Catalogue  (HECATE), a value-added catalogue containing galaxy properties \citep{kovlakas06-21}. 
In Sect.~\ref{sec:samp} we describe the sample selection and source classification. In Sect.~\ref{sec:sixte} we compare observations with the expected detections based on simulations. In Sect.~\ref{sec:scal}, we study the normal galaxy scaling relations based on our results. In Sect. \ref{sec:disc} we discuss the implications of our results. We summarize our conclusions in Sect. \ref{sec:conc}. Throughout the paper, we assume the \citet{kroupa04-01} initial mass function (IMF) for all measured SFRs and stellar masses, and we adjust values from comparison studies where required to conform to a Kroupa IMF. We adopt $\Lambda$CDM cosmology with the following parameters: $H_0=70\,\mathrm{Mpc}\,\mathrm{km}\,\mathrm{s}^{-1}$, $\Omega_\mathrm{M}=0.3$, $\Omega_{\Lambda}=0.7$. 
For fluxes and luminosities we assumed a power-law spectral model with $\Gamma=1.7$ and median Galactic absorption in the direction of the eFEDS field of \nh=\ten{3.6}{20}\,cm$^{-2}$. Unless otherwise stated, uncertainties correspond to 90\% confidence levels.

\begin{figure*}
\centering
\includegraphics[width=1.0\textwidth]{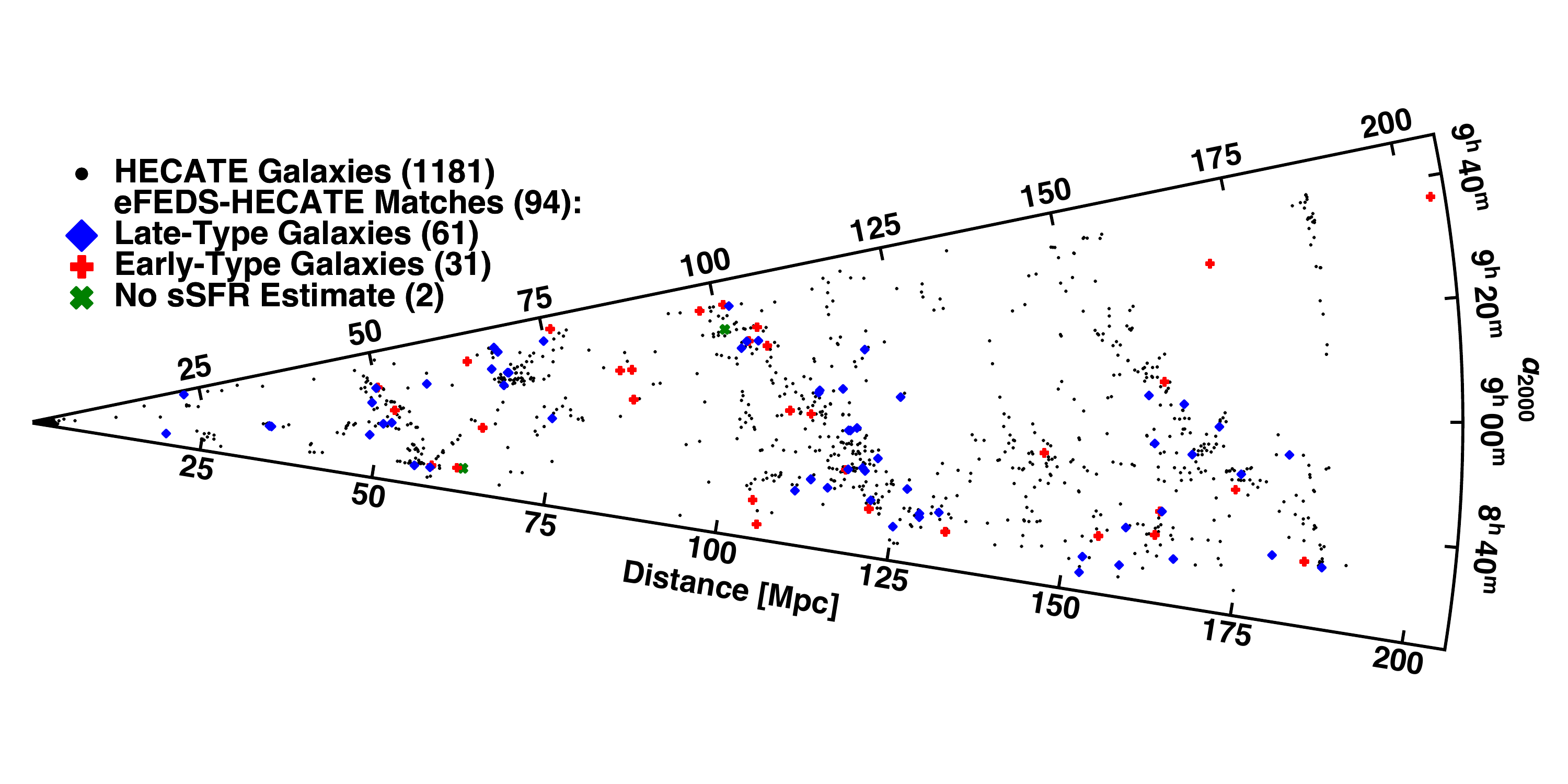}

\caption{Cosmological slice showing the distribution of HECATE galaxies in the eFEDS field (black points). The eFEDS-HECATE matches are subdivided into early-type (red plusses) and late-type (blue diamonds) galaxies, which have been classified solely based on a cut in sSFR, and does not necessarily indicate the nature of the X-ray source emission (e.g., normal star-forming/elliptical galaxy, AGN). Those maches without an estimate of sSFR are shown as green crosses.} 
\label{fig:cosmodist}
\end{figure*}

\section{Sample Selection} \label{sec:samp}

\subsection{eFEDS X-ray Sources} \label{sec:exrs}
The eFEDS field was part of the \ero\ PV phase  observations in 2019 November. eFEDS targeted the Galaxy and Mass Assembly (GAMA) Survey field G09, centred on a right ascension of 136\degr\ ($\sim$9\,h) and declination of $1\fdg4$. Four individual rectangular raster-scan fields of $\sim$35 deg$^{2}$ each were completed for a total exposure of $\approx$360\,ks over \fov\,deg$^{2}$. The average exposure time per source was consistent with that expected for eRASS:8, ${\sim}2.2$\,ks. The scans were completed in survey (scanning) mode, which provided a 26\arcsec\ and 40\arcsec\ average point spread function (PSF) in the 0.5--2\,keV (soft) and 2--10\,keV (hard) bands, respectively, over the 0.83 deg$^{2}$ \ero\ FOV. The X-ray point source sensitivity of eFEDS reached $\sim$\ten{7}{-15}\,\esc\ in the soft band (beyond the final depth of eRASS:8, \ten{1.1}{-14}\,\esc) and $\sim$\ten{2}{-13}\,\esc\ in the hard band (Brunner et al. 2021, submitted). 
From the \nefeds X-ray sources (point-like and extended) in eFEDS detected in the 0.2--2.3\,keV energy band with detection likelihood $\ge5$ (Brunner et al. 2021, submitted), \nefedspe with positional error less than 20\arcsec\ were used as input for matching to galaxies. 
eFEDS astrometry was corrected using the GAIA-unWISE AGN catalogue of \cite{shu11-19}, resulting in astrometric precision of ${\approx}5$\arcsec\ in the eFEDS X-ray catalogue, depending on the number of counts and energy band. 
For a more detailed description of eFEDS and the data reduction procedures, please see the catalogue paper (Brunner et al. 2021, submitted). In addition, the eFEDS X-ray source catalogue provides multiwavelength  photometry from the DESI Legacy Imaging Survey DR8, for the counterparts that were identified in Salvato et al. (2021, submitted), using a combination of NWAY \citep{salvato02-18} paired with machine learning and \texttt{ASTROMATCH} \citep{ruiz10-18}.

\subsection{Galaxy Catalogue} \label{sec:galc}
We used the HECATE value-added galaxy catalogue compiled by \citet{kovlakas06-21}, which includes all galaxies (object type `G') within $D\lesssim200$\,Mpc ($z\lesssim0.045$) from the HyperLEDA database \citep{makarov10-14}. The object type (G) excludes regions in intracluster/group environments, ensuring we do not include galaxies with X-ray emission that might be dominated by hot gas from the intracluster/group environment instead of intragalaxy ISM/XRBs. HECATE also provides positions, sizes, morphological classifications, redshifts, distances, star formation rates (SFRs), stellar masses, and nuclear activity classifications, where available in the literature. In addition, all HECATE galaxies have coordinates with astrometric precision of $<$10\arcsec, and $\sim$92\% have precisions of $<$1\arcsec. For details regarding value-added parameter estimates please see \citet{kovlakas06-21}. By cross-correlating the HECATE catalog with the eFEDS footprint, we identified \hecgal\ galaxies. Due to the incompleteness of multiwavelength parameters, only \hgalssfr of these galaxies have both SFR and stellar mass estimates from HECATE.

\subsection{Catalogue Matching and Source Classification} \label{sec:mclass}
We cross-matched the eFEDS X-ray source and HECATE galaxy catalogues using \texttt{STILTS} \citep{taylor07-06}, which is a tool to process tabular data. Any eFEDS source within the \dtf\ ellipse of a HECATE galaxy was matched using the $\texttt{matcher=skyellipse}$ method. The positional uncertainties for eFEDS X-ray sources (mean of 4.7\arcsec\ for all sources in eFEDS) were included during matching. This yielded 101 matches with a maximum separation of 47\arcsec. Of these, one X-ray source (eFEDS ID 11246) was matched to two unique galaxies (PGC~24658/4532355). We removed one HECATE galaxy matched to eFEDS ID 24454 that was a star-forming region within the main host galaxy that was matched (PGC 25986). We also removed one X-ray source match (eFEDS ID 1286) to PGC 24129 because it was located on the edge of the \dtf\ (likely background AGN), and a better match (eFEDS ID 1284) was coincident with the centre of the galaxy. Finally, we removed five matches that were coincident with foreground stars (confirmed via SDSS spectroscopy and visual inspection), yielding \nmat total matches. From these matches, 3 eFEDS X-ray sources were extended based on their extent and extent likelihood values. We clarify that the eFEDS-HECATE matches represent eFEDS X-ray sources associated with HECATE galaxies, and do not necessarily represent the unique counterpart (if there is a single point source counterpart as opposed to extended emission derived from multiple X-ray sources in the galaxy), which may or may not be part of the galaxy. For eFEDS X-ray sources with  detection likelihoods $\geq$6, there are 7 eFEDS-HECATE matches (eFEDS ID's 2671, 7551, 12847, 17437, 20952, 22198, 29989) for which the HECATE galaxy does not coincide with the counterpart to the eFEDS X-ray source that was identified in Salvato et al. (2021, submitted). In these 7 cases, the centroid of the counterpart assigned in Salvato et al. (2021, submitted) is up to 49\arcsec\ (median of 22\arcsec) from the centre of the HECATE source but still within the \dtf\ ellipse. It is unclear yet whether the X-ray emission is from X-ray emitters within the HECATE galaxies or sources in the background (see Fig. A.1 of Salvato et al. 2021, submitted). In 5/7 of these cases, the match assigned by Salvato et al. (2021, submitted) is coincident with a spiral arm/structure of the HECATE galaxy, indicating a possible ultraluminous X-ray source (ULX) origin. Follow-up with \chandra\ and other high-resolution multiwavelength observations will help localize the X-ray sources and constrain the counterpart(s).

\begin{figure*}
\begin{center}
\begin{tabular}{cc}
\includegraphics[width=0.5\textwidth]{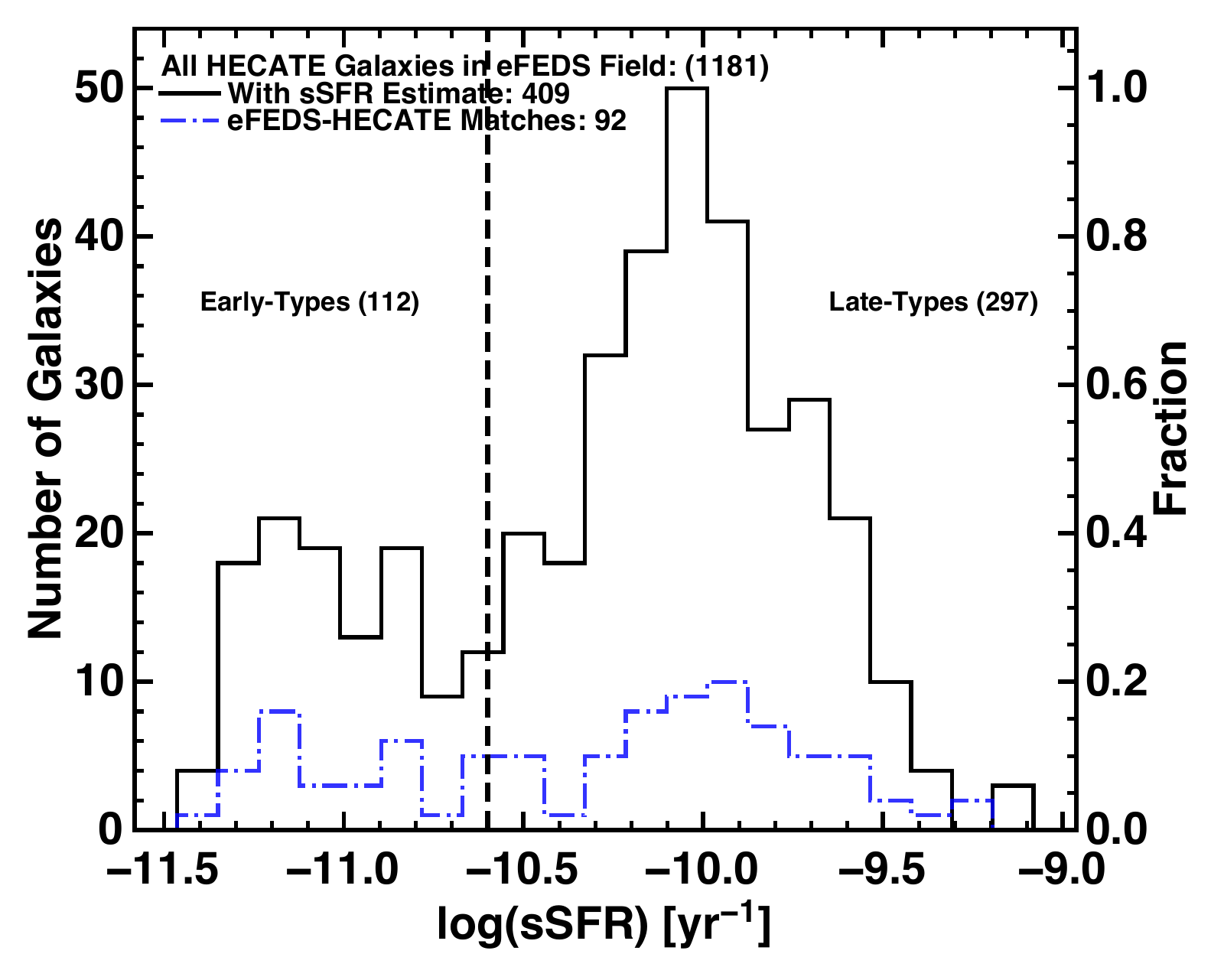}
\includegraphics[width=0.5\textwidth]{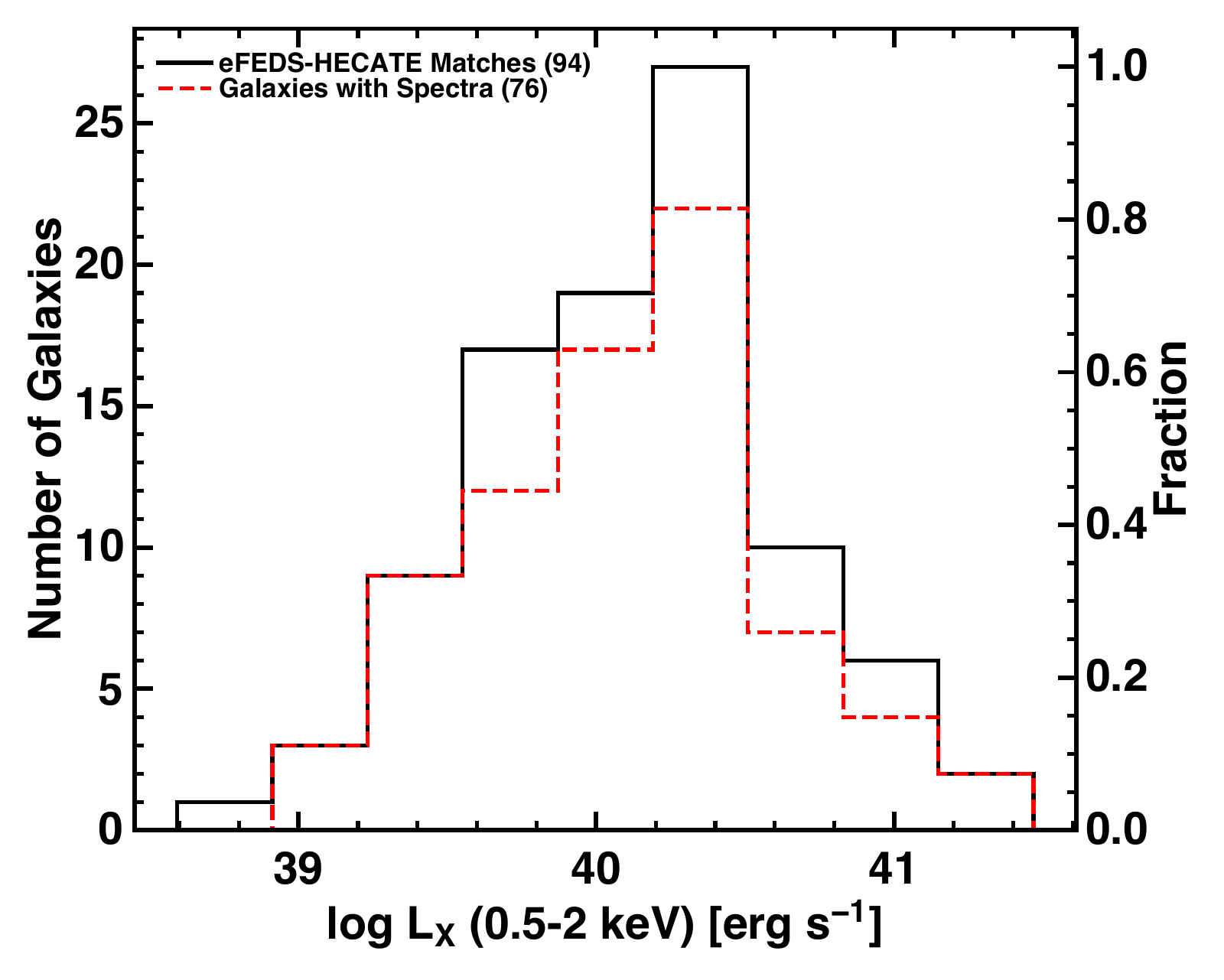}
\end{tabular}
\caption{ \textit{Left:} Histogram of sSFR for all HECATE galaxies in eFEDS with sSFR estimates (solid black line) and those matched to eFEDS X-ray sources (dot-dashed blue line). The bimodal distribution of galaxies is apparent, where $\log \mathrm{sSFR} = -10.6$ (black dashed line) falls between the two peaks and separates the galaxy population into early-types ($< -10.6$) and late-types ($> -10.6$). \textit{Right:} Histogram of \ero\ \lx\ for eFEDS-HECATE matches (solid black line) and those with SDSS/6dF optical spectra (dashed red line), showing that the spectroscopic sub-sample is representative of the eFEDS-HECATE sample. 
{\label{fig:histlxssfr}}%
}
\end{center}
\end{figure*}

Using the value-added parameters from HECATE we can compare the properties of HECATE galaxies in eFEDS to those matched to an eFEDS X-ray source. In Fig.~\ref{fig:cosmodist} we show the cosmological slice representing the eFEDS field, which indicates all HECATE galaxies, the eFEDS-HECATE matches, and early/late-type galaxies (classified only based on a sSFR cut). 
The left panel of Fig.~\ref{fig:histlxssfr} shows histograms of HECATE galaxies in the eFEDS field with specific SFR (sSFR) estimates (black) and those matched to eFEDS X-ray sources (blue dot-dashed). We also indicate the approximate split between late-type and early-type galaxies at $\log \mathrm{sSFR} = -10.6\ \mathrm{yr}^{-1}$ as in \citet{basu-zych08-20}. Based on this adopted value, the sample is dominated by late-type galaxies (73\%), approximately reflecting the cosmological distribution of galaxy types, given the typical stellar mass of the HECATE galaxies in the eFEDS field. For more massive galaxies the fraction of ellipticals would be higher.  

We compared between \ero-detected sources and available archival data to assess variability. Flux calibration of \ero\ on \srg\ was reported in \citet{dennerl12-20}. To investigate archival X-ray data, we cross-matched our eFEDS-HECATE catalogue X-ray positions with the \chandra\ and \xmmn\ source catalogues. 
We found 6 \chandra\ X-ray sources that matched to 5 unique eFEDS X-ray sources (eFEDS ID 150 had two matches within $\sim2$\arcsec) using the 317\,167 unique compact and extended X-ray sources from the \chandra\ Source Catalogue Release 2.0 \citep{evans01-20}. These matches were all within 10\arcsec, consistent with the \ero\ astrometric uncertainty. The 0.5--2 keV fluxes were broadly consistent, showing variability of a factor of two at most. 
The cross-match with the 575\,158 unique X-ray sources in 4XMM-DR10 \citep{webb09-20} resulted in only one match having 6\arcsec\ separation. The flux measurements for this match (eFEDS ID 13787) were consistent. A comparison with the \xmmn\ slew survey source catalogue (XMMSL2) resulted in no matches. The handful of targeted \chandra\ and \xmmn\ observations are short exposures and cover a small fraction of the eFEDS footprint, resulting in the low number of matches. Completion of subsequent eRASS surveys will shed light on the variability (or lack thereof) from normal galaxies. 

To study normal galaxies in the eFEDS field, we require robust optical spectroscopic classifications for each galaxy. This was achieved using optical spectra from the SDSS DR8 MPA-JHU catalogue \citep{kauffmann12-03, brinchmann07-04, tremonti10-04}, the most recent data release with starlight-subtracted spectra. In particular, identifying any type of nuclear (e.g., AGN) activity was critical for assembling the normal galaxy sample. This enabled us to properly investigate normal galaxy scaling relations based only on the X-ray emission from XRBs and hot gas. 
We cross-matched each eFEDS-HECATE X-ray position (using the initial 101 matches) with the nearest SDSS DR8 spectrum within 3\arcmin. This resulted in \nsdss\ matched SDSS spectra. The spectra were all coincident with HECATE galaxy positions except for two that were slightly offset from the galaxy centre by a few arcsec. We initially used the large matching radius of 3\arcmin\ to encompass all nearby spectra and avoid missing spectra coincident with X-ray sources that were offset from the centres of galaxies. We completed visual inspection of each matched spectrum to confirm that the offset from the centre of the HECATE galaxy was $<3\arcsec$, except in 2 cases (eFEDS ID 3372 and 22226) where the spectrum was coincident with a star-forming region. The SDSS \textit{spectrotype} for \nsdssgal/\nsdss\ were GALAXY, with the remaining two being STAR and QSO. After removing the sources described above as foreground stars/duplicate that had spectra, we obtained a total of 57 optical spectra for the sample of \nmat sources. To classify galaxies, we used the SDSS DR8 \textit{subclass} identification in combination with the emission-line ratio classification scheme of \citet{stampoulis05-19}.

\begin{figure}
\begin{center}
\includegraphics[width=1.0\columnwidth]{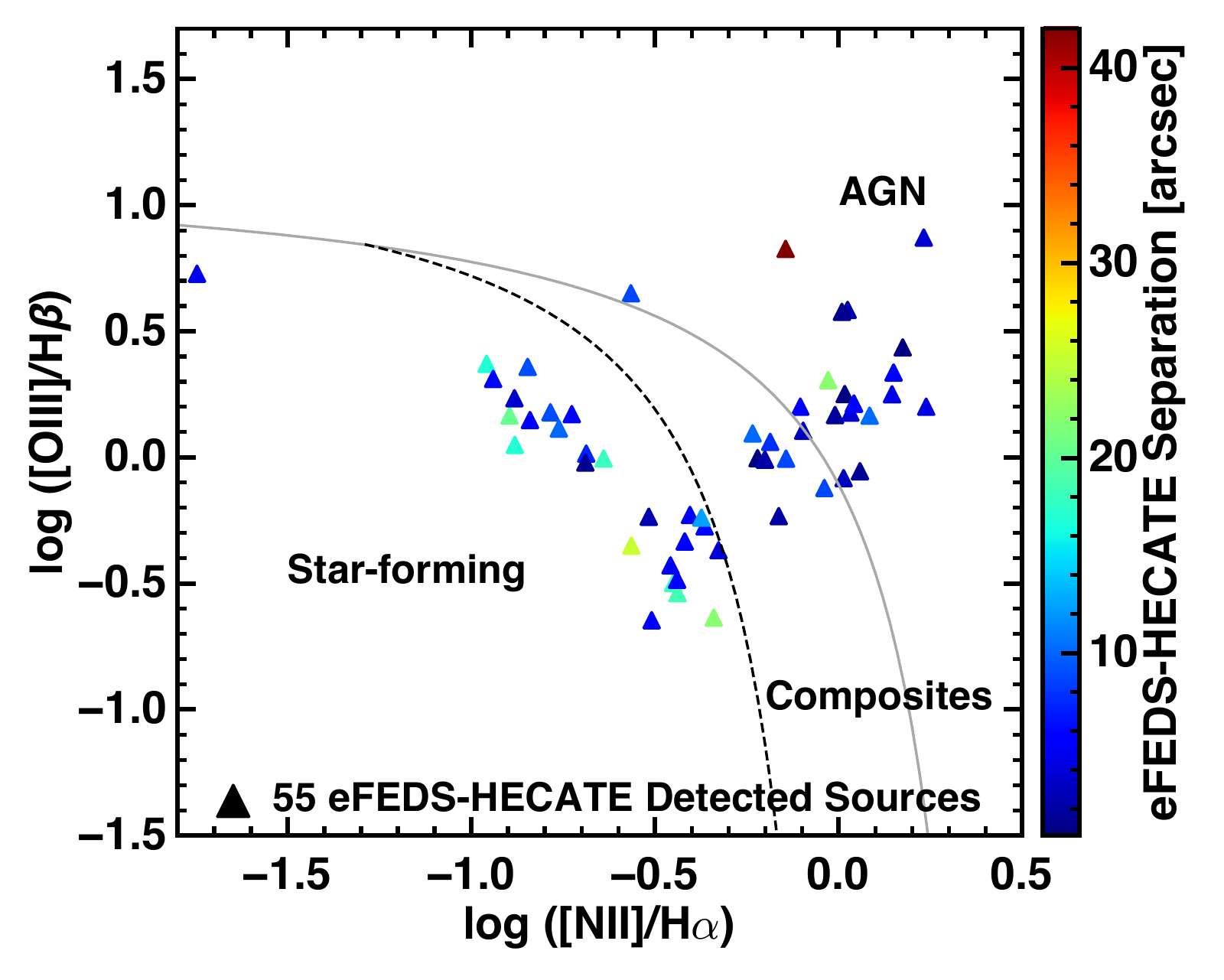}
\caption{BPT diagram, which uses the $\log [\ion{O}{iii}]/\mathrm{H}\beta$ and $\log [\ion{N}{ii}]/\mathrm{H}\alpha$ optical emission line ratios to separate sources into star-forming galaxies, composites, and AGN classes. To obtain robust classifications, we also incorporated $\log [\ion{S}{ii}]/\mathrm{H}\alpha$ and $\log [\ion{O}{i}]/\mathrm{H}\alpha$ line ratios following \citet{stampoulis05-19}. The colorbar indicates the separation between the eFEDS X-ray source and the HECATE galaxy in arcsec. 
{\label{fig:bpt}}%
}
\end{center}
\end{figure}

Galaxy classification is critical to separate normal galaxies from those with nuclear activity due to SMBH accretion. For example, the use of Baldwin-Phillips-Terlevich (BPT) diagrams \citep{baldwin02-81} has become ubiquitous to separate star-forming galaxies from galaxies dominated by AGN activity, and later included Low-Ionization Nuclear Emission Regions (LINERs). However, the delineation between these different source classes, which is defined using theoretical and empirical curves, is subject to considerable uncertainty, where X-ray detected AGN have been found in the star-forming region \citep{agostino05-19, lamassa05-19}. In Fig.\,\ref{fig:bpt} we show the BPT diagram for the \nsdssgal\ eFEDS-HECATE sources with SDSS spectra, of which half fall in the star-forming region based on the classification curves. To obtain the probabilities of each classification type (star-forming galaxy, Seyfert, LINER, and composite) for each source we used the code provided by \citet{stampoulis05-19} and their clustering prescriptions from their Table~\ref{tab:mcres}. Galaxies were classified using the optical emission line ratios $\log [\ion{N}{ii}]/\mathrm{H}\alpha$, $\log [\ion{S}{ii}]/\mathrm{H}\alpha$, $\log [\ion{O}{i}]/\mathrm{H}\alpha$, and $\log [\ion{O}{iii}]/\mathrm{H}\beta$. For the purposes of the present work, we combined Seyfert and LINER classifications into the AGN category. In the case that the probabilities prevent a classification, the result is undefined. For half of these undefined sources, their SDSS DR8 \textit{subclass} identification was BROADLINE, which we adopted. The \citet{stampoulis05-19} methodology does not account for broad lines, and thus was unable to classify these sources as such.

To add additional optical spectroscopic classifications, we cross-matched the positions of the remaining eFEDS-HECATE sources with the Final Release of the 6dF Galaxy Survey \citep{jones12-04, jones10-09} by taking the nearest spectrum within a circle of radius 7\arcsec. This resulted in 24 additional galaxy spectra. After we de-redshifted the spectra, we removed the starlight component by using the \texttt{STARLIGHT} code \citep{cid-fernandes04-05, mateus08-06}. We then classified them into star-forming, composite, and AGN by visually inspecting the spectra. We identified three objects with broad Balmer lines which were clearly classified as AGN. For the remaining objects, we estimated the line ratios of $[\ion{O}{iii}]/\mathrm{H}\beta$ and $[\ion{N}{ii}]/\mathrm{H}\alpha$ lines. A comparison with the line-ratio diagnostics of \citet{stampoulis05-19} suggested that 9 of them were star-forming galaxies and three were composites. Because of issues with the flux calibration of the 6dF spectra, these classifications refer only to objects well within their respective loci in the $[\ion{O}{iii}]/\mathrm{H}\beta$--$[\ion{N}{ii}]/\mathrm{H}\alpha$ diagnostic diagram. Furthermore, we characterized three additional galaxies as star-forming/composites because it was not clear in which class they belong (we adopted the composite classification for these 3), while for five galaxies the quality of the spectrum was inadequate for any classification. Finally, the spectrum of one galaxy was dominated by strong absorption lines and as a result, we classified it as an elliptical galaxy. Despite the absence of accurate measurements for the full sample of line ratios used in galaxy activity diagnostics, the visual examination of the spectra showed that the obtained classifications are robust. As a result, we incorporated an additional 19 classifications from the eFEDS-HECATE sample, which included 10 normal galaxies. 
Therefore we obtained spectroscopic classifications for \nspec\ galaxies, with \normgal classified as normal galaxies.

In the right panel of Fig.~\ref{fig:histlxssfr} we show histograms of \lx\ for all eFEDS-HECATE matches and those with optical spectra. In Fig.~\ref{fig:histlxssfrclass} we show the same plots from Fig.~\ref{fig:histlxssfr} that are now divided into the classifications summarized above. The `undefined' class (red dot-dot-dashed lines) represents sources with optical spectra that were not able to be classified, as opposed to eFEDS-HECATE sources with no optical spectra and thus no classification. In the left panel, the separation between star-forming galaxies in the late-type region vs.\ AGN in the early-type region is evident. This separation is due to the selection via emission lines, which are produced by strongly ionizing sources such as either hot, massive stars (i.e., star-forming galaxies) or AGN. The right panel shows that the AGN have higher \lx\ compared with star-forming galaxies, as expected. It is important to consider that applying optical emission line diagnostics biases this work towards the selection of star-forming galaxies and omits early-type galaxies, hence the dearth of classifications for the latter. 

To further probe the nature of the eFEDS-HECATE galaxies, we decompose sSFR and plot the SFR versus stellar mass in Fig.~\ref{fig:sfrmass}. All HECATE galaxies in the eFEDS field are shown as black dots. We indicate the source classifications determined for \nspec\ sources via spectroscopy using various unfilled symbols and colours. 
We show star-forming galaxies (blue triangles), composite galaxies (green circles), AGN (orange triangles), sources with spectra that could not be classified and are thus undefined (red crosses), and sources with no spectroscopic information (magenta stars). There is a delineation between star-forming galaxies that are found at high-sSFR and the AGN, which are preferentially hosted in early-type galaxies, at low-sSFR. 
In Table~\ref{tab:mcres} we summarize the results of eFEDS-HECATE matching and the source classifications.

\subsection{Purity of the eFEDS-HECATE Normal Galaxies}

We assessed the purity of the sample of classified normal galaxies using a number of different methods. First, we determined the false-match rate via Monte Carlo simulations. We created 10\,000 galaxy catalogues, each with the same number of galaxies (\hecgal) as the original matched HECATE catalogue. We matched each of these catalogues to the eFEDS X-ray source catalogue and derived a false-match rate of 1.5\%, corresponding to $\approx0.6$ false sources. 
We also checked for the incidence of background AGN within the \dtf\ ellipses of all eFEDS-HECATE matches that were classified as normal galaxies in Section \ref{sec:mclass}. We first compared with the number counts from the CDF-S 7 Ms Survey \citep{luo01-17} because of the completeness at and well below the eFEDS sensitivity limit of $\sim$\ten{7}{-15}\,\esc\ in the 0.5--2\,keV energy band. We utilized the cumulative number of AGN per deg$^{2}$ expected at and above this limit and multiplied by the total area within the \dtf\ ellipses of all eFEDS-HECATE matches that were classified as normal galaxies. This resulted in only 0.4 expected background AGN. We obtained the same value using the number counts from the larger-area, 9.3 deg$^{2}$ \chandra\ survey of the Bo\"otes Field \citep{masini11-20}. The number counts from the \xmmn\ 2 deg$^{2}$ survey of the COSMOS field \citep{cappelluti04-09} yielded a slightly larger estimate of 0.45 expected background AGN. When using the area within the \dtf\ of all \nmat\ eFEDS-HECATE matches, which includes those that were classified as AGN-hosting, we obtained 1.4 expected background AGN. 

We also assessed the likelihood that the eFEDS-HECATE normal galaxies were spurious using the results of source detection simulations completed by Liu et al. (2021, submitted). Using the spurious source fraction and the distribution of detection likelihood for the eFEDS-HECATE normal galaxies, we estimated 2 spurious sources. Taken together with the estimates above, the normal galaxy sample potentially has $\sim3$ sources of questionable origin, but is otherwise well-determined.

\subsection{Metallicity Estimates} \label{sec:metal}
We obtained galaxy gas-phase metallicities for the eFEDS-HECATE sample using the SDSS DR8 MPA-JHU catalogue starlight-subtracted optical emission line flux measurements \citep{brinchmann07-04}. The catalogue used stellar population synthesis models to fit and subtract the stellar continuum, including stellar absorption features, enabling the accurate measurement of nebular emission lines. Following the O3N2 method of \citet[][equation 3]{pettini03-04}, we calculated $12+\log (\rm{O/H})$ using the $[\ion{O}{iii}]$, $[\ion{N}{ii}]$, $\mathrm{H}\alpha$, and $\mathrm{H}\beta$ emission line fluxes. \citet{kewley07-08} showed the O3N2 method was able to trace a wide range of metallicities, had relatively low scatter, and was comparatively less sensitive to extinction effects. We calculated metallicities for all \nsdssgal\ eFEDS-HECATE/SDSS matched sources that had the SDSS \textit{spectrotype}$=$GALAXY designation. This includes metallicities for \metng\ star-forming galaxies. The metallicity estimates from the \citet{pettini03-04} empirical relation have systematic uncertainties of $\pm0.4$\,dex, which dominate when compared with statistical uncertainties. In Table \ref{tab:srcchar}, we present the source characteristics for each eFEDS-HECATE match.

\begin{table}
\caption{eFEDS-HECATE Matching and Classification Results}         
\label{tab:mcres}   
\centering              
\begin{tabular}{@{}lll@{}}         
\hline\noalign{\smallskip}               
Description & Number	&	N(sSFR)\tablefootmark{a}	\\    
\noalign{\smallskip}\hline\noalign{\smallskip}
HECATE Galaxies in eFEDS Field	&	\hecgal	&	\hgalssfr	\\
eFEDS X-ray Sources         	&	\nefedspe	&		\\
eFEDS-HECATE Matches		&	\nmat	&	92	\\
eFEDS-HECATE Matches with Spectra	&	76	&	75	\\
\noalign{\smallskip}\hline\noalign{\smallskip}
\multicolumn{3}{c}{Classifications}	\\
\hline\noalign{\smallskip}
Star-forming	&	\sfgal	&	35	\\
Early-type		&	\egal	&	1	\\
Composite   	&	11	&	11	\\
AGN			    &	24	&	24	\\
Undefined		&	4	&	4	\\
No Spectrum  	&	18	&	17	\\

\hline                                            
\end{tabular}
\tablefoot{sSFR estimates are unavailable for some HECATE galaxies, and thus the total number that appears in figures using sSFR will vary. 
\tablefoottext{a}{Number that have an sSFR estimate.}}
\end{table}

\begin{figure*}
\begin{center}
\begin{tabular}{cc}
\includegraphics[width=0.5\textwidth]{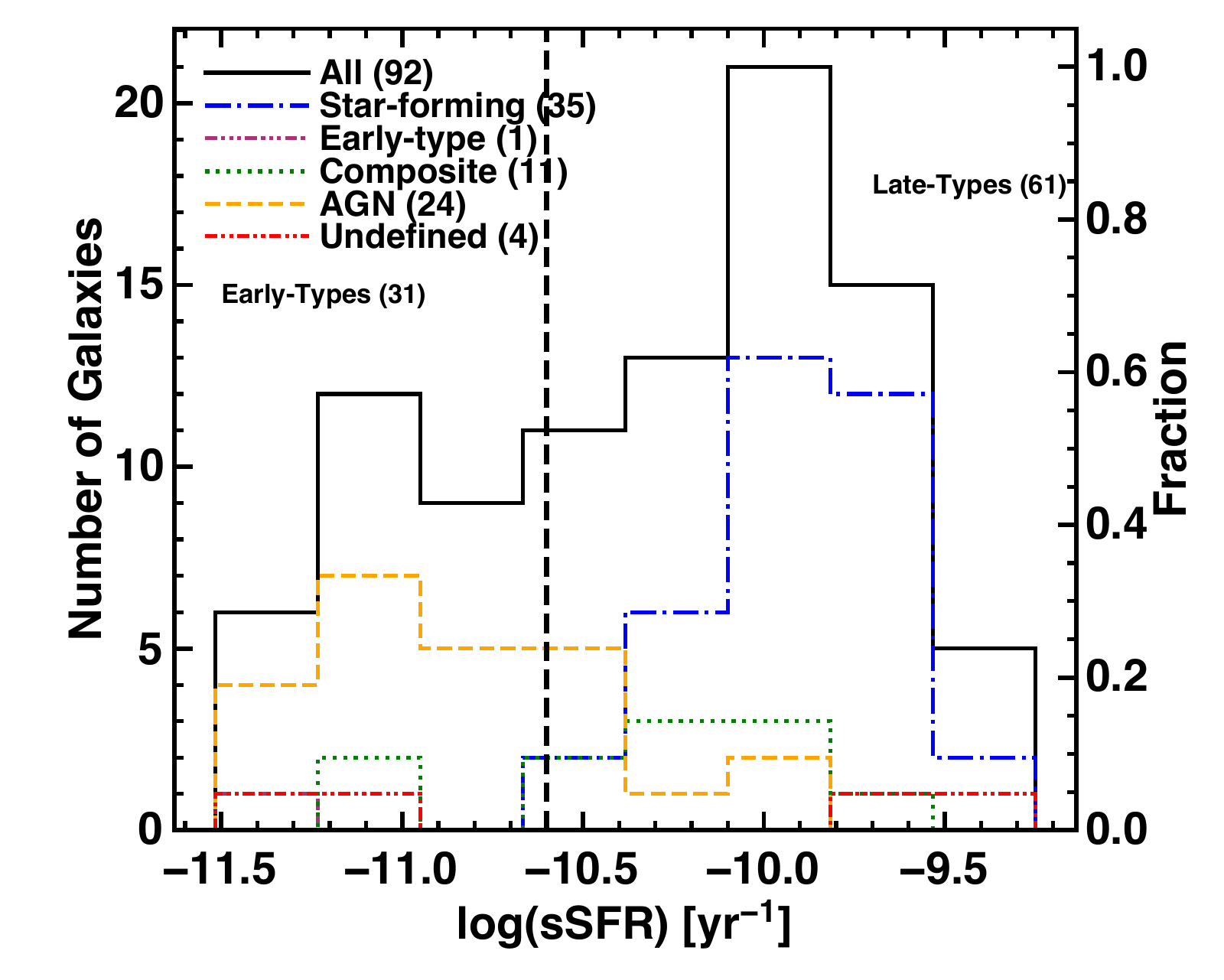}
\includegraphics[width=0.5\textwidth]{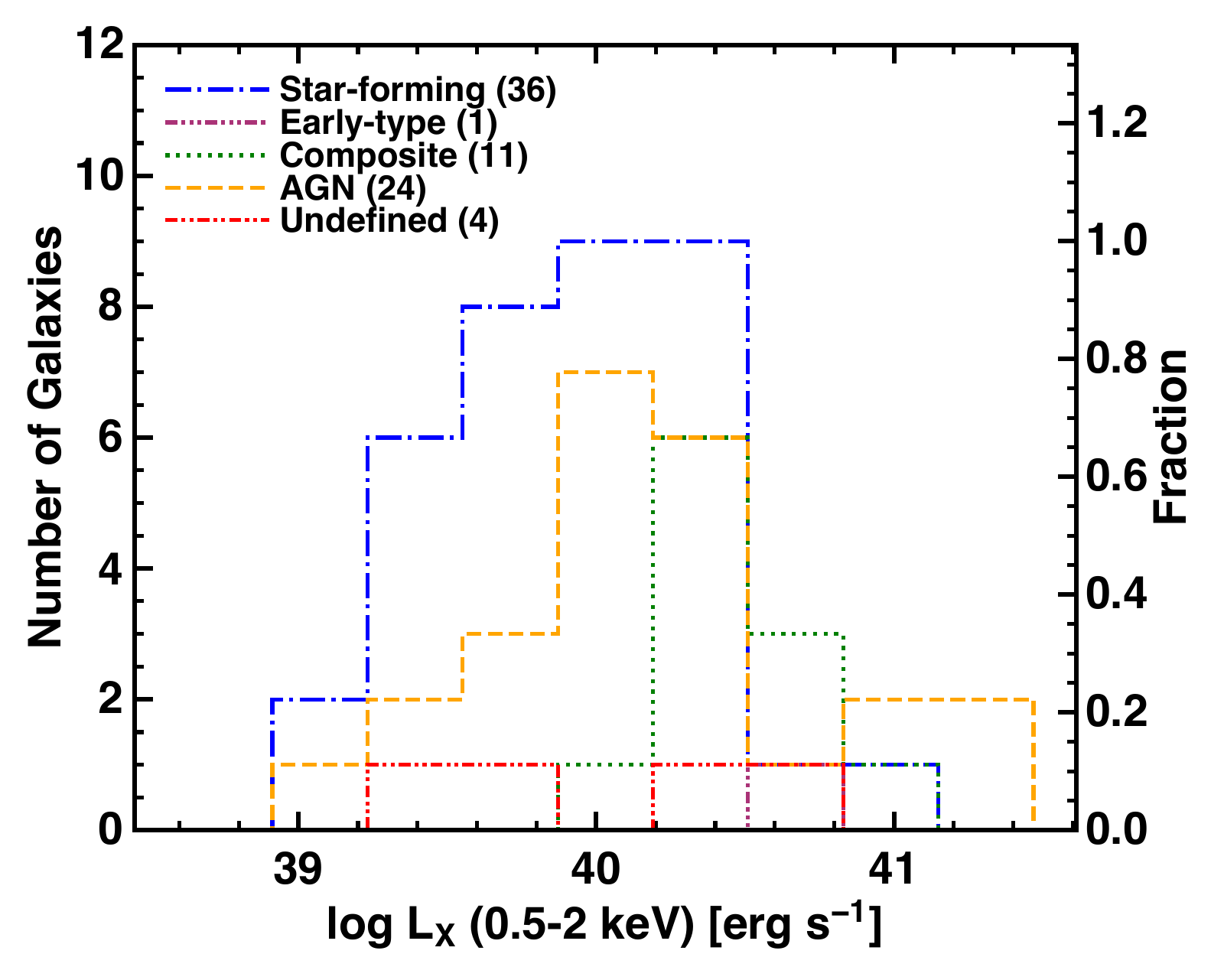}
\end{tabular}
\caption{\textit{Left:} Histograms showing sSFR of eFEDS-HECATE matches (black) and the sub-classifications based on sources matched to SDSS/6dF optical spectra. Details of the classification methods are described in Sect. \ref{sec:mclass}. \textit{Right:} Histograms of \ero\ \lx\ showing the sub-classifications based on SDSS/6dF optical spectroscopy of eFEDS-HECATE matches. The remaining eFEDS-HECATE matches are missing optical spectroscopic data and thus do not have sub-classifications. 
{\label{fig:histlxssfrclass}}%
}
\end{center}
\end{figure*}

\begin{figure}
\begin{center}
\includegraphics[width=0.5\textwidth]{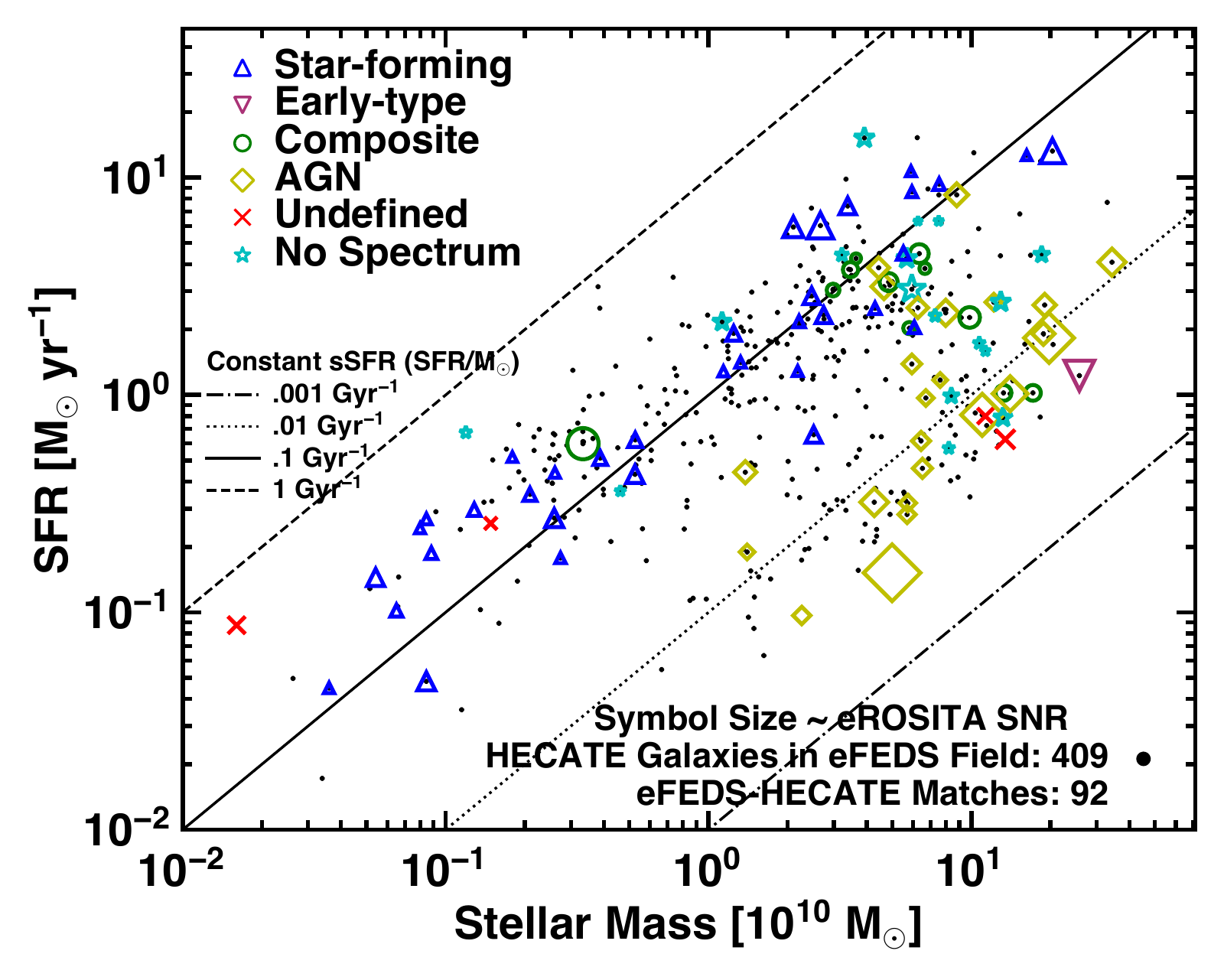}
\caption{SFRs and stellar masses $M_{\star}$ for all HECATE galaxies in the eFEDS field (black points) and the eFEDS-HECATE matches split into their respective classifications as described in Sect. \ref{sec:mclass}. Only galaxies that have these parameter estimates are shown. The symbol size is proportional to the \ero\ detection likelihood. We also overlay lines of sSFR for reference. The AGN-hosting galaxies are at preferentially lower SFR (early-type) when compared with the star-forming galaxies. We only plotted galaxies having \mstar\ $>10^{8}$ \msun; there are 13 galaxies below this limit. 
{\label{fig:sfrmass}}%
}
\end{center}
\end{figure}

\section{\ero\ Observations vs. Simulations}\label{sec:sixte}

In \citet{basu-zych08-20}, predictions were made for eRASS detections of normal galaxies based on the HECATE galaxy catalogue and prescriptions for the X-ray emission from normal galaxies. Simulations were completed both analytically and using the Simulation of X-ray Telescopes software \citep[SIXTE,][]{dauser09-19}. In this section we compare the predictions for eRASS:8 simulations with eFEDS observations as these surveys have similar sensitivity, albeit eFEDS only covers \fov deg$^{2}$. 

\subsection{Predicted vs.\ Observed Normal Galaxy X-ray Flux}

\begin{figure}
\begin{center}
\includegraphics[width=0.5\textwidth]{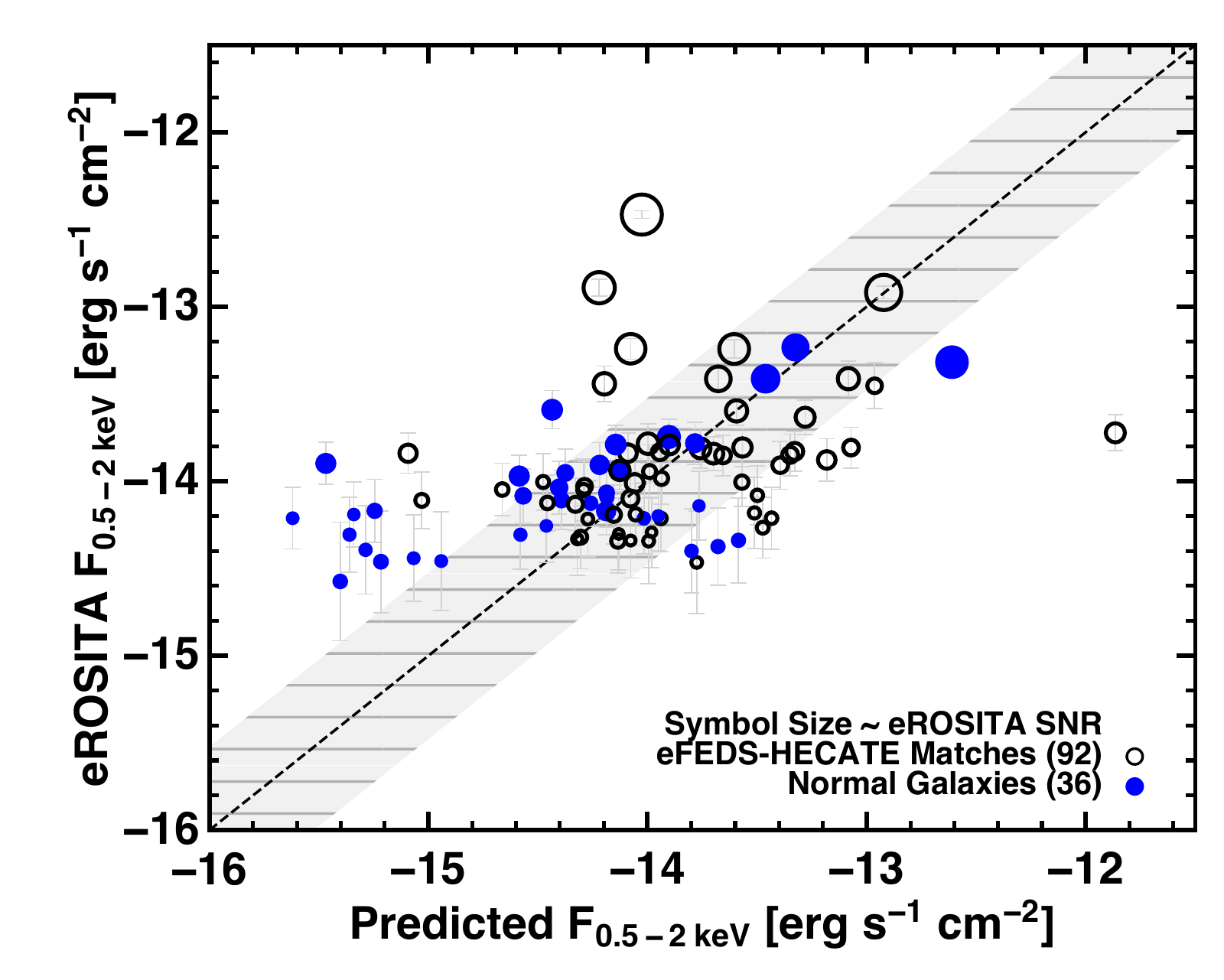}
\vspace{0.in}
\caption{Comparison between the integrated (XRB + hot gas) flux predicted using normal galaxy scaling relations as in \citet{basu-zych08-20} and the observed \ero\ flux from sources detected in the eFEDS detected. All eFEDS-HECATE matches with sSFR estimates are shown as unfilled black circles, whereas the spectroscopically classified normal galaxies are shown as filled blue circles. The light grey area represents the approximate dispersion from the scaling relations used to predict the flux of normal galaxies. \label{fig:sixteflux}}
\end{center}
\end{figure}

In Fig.\,\ref{fig:sixteflux} we show the \ero\ flux for eFEDS sources compared with the predicted flux based on prescriptions for the X-ray emission from normal galaxies (XRBs + hot gas). This predicted flux is different from the flux expected for a source that was detected in SIXTE simulations, due to the adopted sensitivity limit and variations in the background. 
We can only show sources for which we have sSFR estimates such that their X-ray fluxes were accurately predicted (using stellar mass and SFR to estimate low-mass XRB (LMXB), high-mass XRB (HMXB), and hot gas flux). 
For the \normgal spectroscopically classified normal galaxies, there is a large offset in the observed vs.\ predicted flux for sources with predicted fluxes below \ten{5}{-15} \esc, given the sensitivity limit for eFEDS.
Most of the sources with these low predicted fluxes that were detected by \ero\ are likely a result of stochasticity in the sample of normal galaxies, which is a larger effect at lower SFRs and stellar masses \citep{justham06-12, lehmer01-21}. The scatter in the scaling relations for each component of the integrated emission from normal galaxies (LMXBs, HMXBs, and hot gas) was estimated in \citet{basu-zych08-20} to vary based on morphological type and flux (see their Figs. 14 and 15), with dispersion in flux up to a factor of $\sim3$. The variation in predicted vs.\ observed flux for star-forming galaxies at the lowest predicted fluxes in Fig.\,\ref{fig:sixteflux} is outside of this range, indicating additional effects such as low stellar-mass (dwarf galaxies - outside the range of values where scaling relations are derived), stellar age, metallicity, and merger history may likely contribute to the scatter.

\subsection{Characteristics of Simulated \& Observed Samples}	\label{sec:charsixt}

\begin{figure}
\begin{center}
\includegraphics[width=1.0\columnwidth]{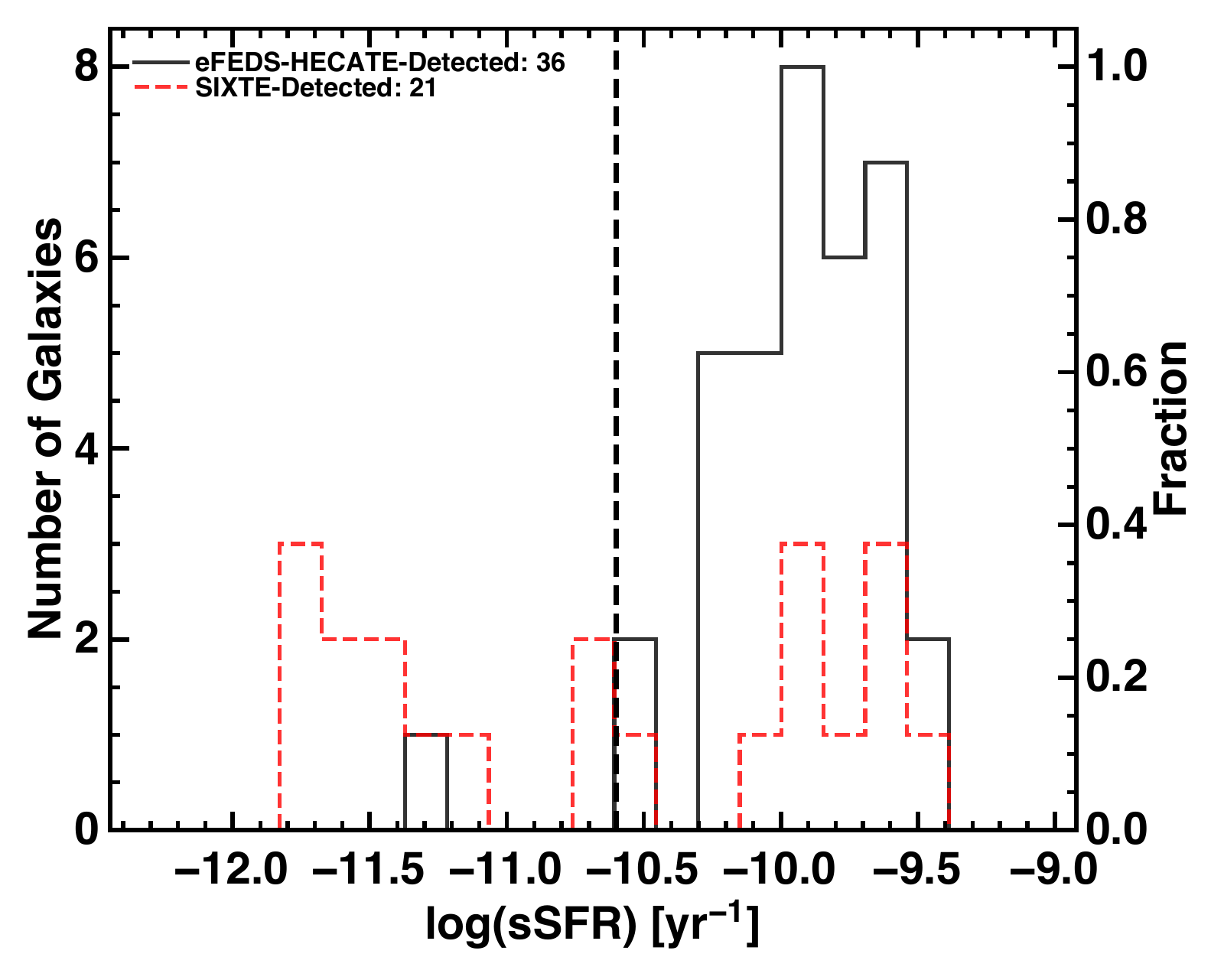}
\caption{Histogram of sSFR comparing eFEDS-HECATE normal galaxies detected in \ero\ observations (solid black line) with the predictions for eFEDS detections based on SIXTE simulations (dashed red line). The black vertical dashed line shows the sSFR cut adopted to separate early-type and late-type galaxies. eFEDS observations detected almost exclusively late-type galaxies, at odds with SIXTE simulations that predicted an approximately even split within the eFEDS field. 
{\label{fig:sixte}}
}
\end{center}
\end{figure}

The SIXTE simulations from \citet{basu-zych08-20} found that most detected normal galaxies peaked at low-sSFR, coincident with the most massive early-type galaxies. A longer tail towards the highest-sSFR late-type galaxies was found (see Fig.\,9 of \citealt{basu-zych08-20}). The eRASS:8 SIXTE simulations predicted 3318 detections of normal galaxies, where 67\% were early-types and 33\% were late-types. Within the eFEDS field, this ratio was markedly different, with 11 (52\%) early-type and 10 (48\%) late-type normal galaxy SIXTE-detected predictions. The HECATE catalogue excludes galaxies residing in group/cluster environments, which happen to host more massive early-type galaxies. However, eRASS:8 simulations also used the HECATE catalogue, pointing to potential effects from cosmic variance. In Fig.\,\ref{fig:sixte} we show a sSFR histogram comparing eFEDS normal galaxies detected in \ero\ observations and SIXTE simulations. While it seems there are approximately twice as many detections from \ero\ observations compared to detections in the SIXTE simulations, there are a number of issues with the characteristics of each sample and their distributions. Firstly, the bias in sSFR is clearly evident for early-type and late-type galaxies. SIXTE simulations showed an approximately even split whereas \ero\ observations almost exclusively identified late-type galaxies (\sfgal/\normgal). Based on SIXTE simulations from \citet{basu-zych08-20}, we predicted 11 early-type normal galaxy detections within eFEDS- so why were more not identified in \ero\ observations? Optical spectroscopy is biased against the selection of early-type normal galaxies because the methodology is based on detecting emission lines, which are prevalent in star-forming galaxies (very hot gas and OB stars) and galaxies hosting AGN. However, elliptical galaxies are characterized by strong absorption lines due to metals in the stellar atmospheres of low-mass stars, but have no strong emission lines. We spectroscopically identified 18 low-luminosity AGN and 3 composites in the early-type galaxy region based on our sSFR cut, whereas an additional 7 galaxies in this region had no spectrum. Of these 7 sources, the morphological classifications are Sa, Sab, Scd, and 4 S0-types. 
If we assumed these 7 galaxies were not detected via optical spectroscopy due to the presence of absorption lines and lack of AGN emission lines, and are thus normal early-type galaxies based on our sSFR cut, \ero\ observations would still not be in agreement with the predictions from SIXTE simulations.

We find agreement between the overall number of detections from \ero\ observations and SIXTE simulations when comparing flux detection and distance limits between each method. Due to variations in the flux sensitivity of both methods, we only compared sources detected above a given sensitivity limit. The SIXTE simulations used a lower limit on distance of 50 Mpc, whereas \ero\ observations included galaxies detected at all distances. Using the flux limit from the SIXTE simulations in the eFEDS field, along with the distance cut of 50 Mpc, we determined there are 22 normal galaxies detected from \ero\ observations, consistent with the 21 normal galaxies detected by SIXTE simulations.

However, the large discrepancy between observations and simulations is revealed when incorporating a key final result from \citet{basu-zych08-20} concerning galaxy distributions. When analysing the SDSS optical spectroscopy available for simulated HECATE `normal' galaxies, they found that only 22.6\% of all detections were comprised of confirmed normal galaxies (XRBs + hot gas), whereas Seyferts (23.4\%), LINERS (13.7\%), and composites (40.2\%) accounted for the remaining detections. Applying this ratio to SIXTE simulations of the eFEDS field would result in only 5 normal galaxy detections, 3 early-type and 2 late-type. This presents a critical issue for late-type galaxies, as there exists a factor $\approx11$ discrepancy between observed and simulated samples. Conversely, the early-type galaxy comparison is somewhat improved because the number of detections from SIXTE simulations has been reduced. In fact, Fig.\,\ref{fig:histlxssfrclass} showed that most classifications of eFEDS-HECATE sources in the early-type region are low-luminosity AGN (Seyfert/LINER). When applying the ratios calculated in \citet{basu-zych08-20}, we assumed that there is no bias between morphological type. However, the majority of galaxies detected in simulations were early-types, which are skewed towards having AGN-like optical spectra, meaning that a larger fraction were identified as AGN in the ratios presented above. Conversely, most late-type galaxies detected in simulations had star-forming optical spectra, and thus preferentially classified as normal galaxies. If we approximated and accounted for the late/early-type designation in applying the ratios above, the discrepancy between late-types in simulations vs. observations would only be a factor of a few as opposed to $\approx11$. One caveat to these estimates is that there are 18 eFEDS-HECATE matches that have no spectrum and 4 that have a spectrum that cannot be classified, indicating the potential for an additional 22 normal galaxy detections, approximately split evenly into early-type and late-type.

Not only are the predictions from SIXTE simulations inconsistent with \ero\ observations, but the intersection between each sample of detected galaxies is minimal -- only 3 galaxies were detected in both the \ero\ observations and the SIXTE simulations. The most concerning aspect of this result is that the prescriptions used for estimating X-ray emission of normal galaxies were not successful in predicting which galaxies would be detected. This points to an incomplete understanding of the X-ray emission from normal galaxy populations, particularly the hot gas scaling relations and contribution from dwarf galaxies, as well as stellar age, metallicity, merger history, and low-level/obscured nuclear activity.

\subsection{\ero\ All-Sky Survey Estimates Compared}

\citet{basu-zych08-20} predicted that eRASS:8 would detect approximately 15\,000 normal galaxies over the whole sky after the completion of the 4-year survey. This was based on extrapolation from detections in a much smaller area of the sky (see their Fig.\,1), due to the incompleteness of galaxy catalogues, particularly towards the faint end (e.g., dwarf galaxies), and the varying sensitivity limits of galaxy surveys. Nonetheless, these predictions provided a lower limit of the number of normal galaxies \ero\ should detect. However, as discussed in Sect. \ref{sec:charsixt}, a crucial caveat to these expectations was that these `normal' galaxies are composed of a distribution of different supermassive black hole accretion rates, represented by quiescent (\ie normal galaxy), composite, LINERS, and Seyferts. The result is that only 22.6\% of the predicted 15\,000 `normal' galaxy detections were confirmed normal galaxies with quiescent supermassive black holes, whereas the remaining galaxies had some type of low-level AGN activity. Therefore SIXTE simulations predicted eRASS:8 would detect a lower limit of $\approx3400$ confirmed normal galaxies (XRBs + hot gas) after the 4-year all-sky survey.

Using the results of \ero\ observations from the \fov deg$^{2}$ eFEDS field, we estimated the number of normal galaxy detections expected from eRASS:8 because we expect that the sensitivity limit of each survey will be identical. From the \normgal normal galaxies detected by eFEDS observations, we scaled the eFEDS survey area to the area of the whole sky to obtain an estimate of $\approx12\,500$ eRASS:8 normal galaxy detections. This represents the number of detections based on the sensitivity limit from eFEDS observations. To compare directly to the SIXTE simulation prediction of \citet{basu-zych08-20}, we used only 22/\normgal normal galaxies detected by eFEDS observations at the same sensitivity limit that was used for SIXTE simulations, as well as the distance cut of $D\geq50$ Mpc. This one-to-one comparison resulted in an estimate of $\approx6530$ eRASS:8 normal galaxy detections, a factor of 1.9 larger than the $\approx3400$ confirmed normal galaxies expected from SIXTE simulations. 

To further put these results in context, while SIXTE simulations for eRASS:8 underestimated the number of normal galaxies by a factor of 2, there are still 22 eFEDS-observed sources that have no classification and are potential normal galaxies. In addition, there are potentially tens of additional normal galaxies, particularly in the dwarf regime, that are eFEDS-detected but as of yet not identified as galaxies from, e.g., optical surveys. Table \ref{tab:sixtcomp} summarizes the results of comparisons between SIXTE predictions and \ero\ observations.

\begin{table}
\caption{Normal Galaxy Detections: Simulation vs. Observation}
\label{tab:sixtcomp}   
\centering              
\begin{tabular}{@{}lll@{}}         
\noalign{\smallskip}\hline\noalign{\smallskip}
Description & SIXTE Prediction	&	Observed	\\ 
\hline\noalign{\smallskip}
eFEDS	&	21	&	\normgal	\\
eRASS:8	&	15\,000	&		\\
\hline\noalign{\smallskip}
\multicolumn{3}{c}{Confirmed Normal Galaxies\tablefootmark{a}}	\\
\hline\noalign{\smallskip}
eFEDS	&	5	&	\normgal	\\
eRASS:8	&	3400	&	12\,500\tablefootmark{b}	\\
\hline\noalign{\smallskip}
\multicolumn{3}{c}{Identical Flux \& Distance Limits}	\\	\hline\noalign{\smallskip}
eFEDS	&	5	&	22	\\
eRASS:8	&	3400	&	6530\tablefootmark{b}	\\
\hline                                            
\end{tabular}
\tablefoot{
See Sect. \ref{sec:sixte} for details regarding different subclassifications and extrapolation for eRASS:8. \\
\tablefoottext{a}{Normal galaxies confirmed using optical spectroscopy.}
\tablefoottext{b}{Extrapolation based on eFEDS detections.}
}
\end{table}

\section{Comparison With Normal Galaxy Scaling Relations}\label{sec:scal}

\subsection{Scaling with Specific Star Formation Rate}\label{sec:scal1}

The eFEDS-HECATE matches that are classified as normal galaxies (which are all star-forming based on the emission line diagnostics) can be compared with previous normal galaxy studies as a result of the ancillary galaxy parameter measurements that are available (SFR, stellar mass, metallicity). One caveat to the scaling relations studied in this section is that \ero\ detected the integrated emission from these galaxies in the soft band (0.5--2 keV), whereas only 2 normal galaxies (eFEDS IDs 3372 and 29989) were detected in the \ero\ hard band (2--5 keV). Previous work with \chandra\ and \xmmn\ has focused on the resolved XRB emission in galaxies and how that scales with galaxy SFR and stellar mass, but has not been combined with the hot gas emission. \citet{mineo11-12} and \citet{mineo01-142} studied the properties of hot gas and integrated X-ray emission from star-forming galaxies and provided estimates of its contribution based on SFR. We utilized these results below to put our unresolved \ero\ detections in context. 

\begin{figure*}
\begin{center}
\includegraphics[width=1.0\textwidth]{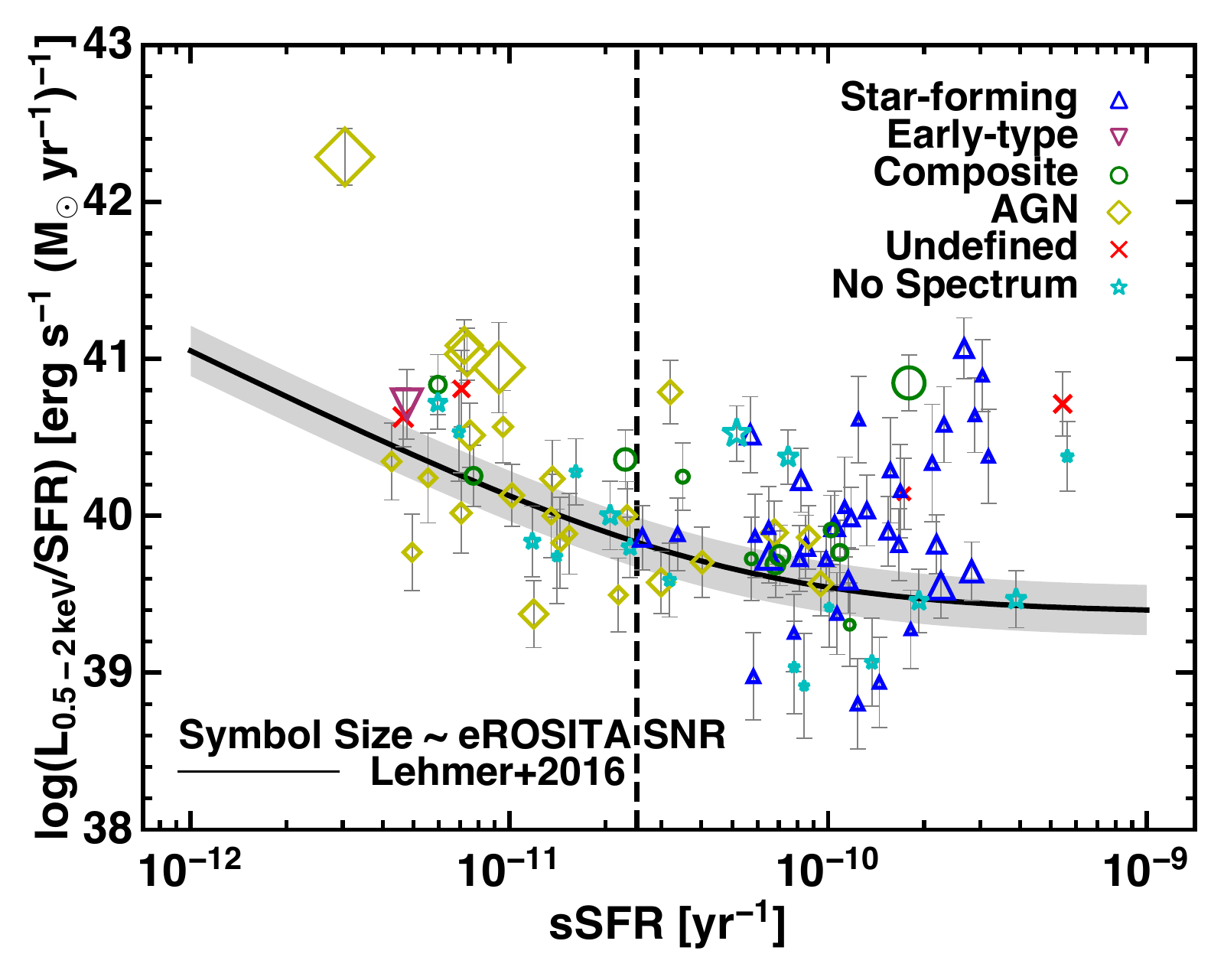}
\caption{Total 0.5--2.0 keV \lx\ scaled by SFR as a function of sSFR for each \ero\ source matched to a HECATE galaxy. The X-ray emission for each source includes that from XRBs and hot gas. The solid black line represents the scaling relation for normal galaxies from \citet{lehmer07-16}, which includes local galaxies and stacked samples from the CDF-S. Source classifications are the same as those from Fig.\,\ref{fig:sfrmass}. Many star-forming galaxies, especially those at high-sSFR, show large dispersion about the scaling relation. The AGN and composite sources are broadly consistent with the scaling relation, which is unexpected, indicating obscured and/or low-level AGN emission. The handful of star-forming galaxies at high-sSFR that are offset from the relation are dwarf galaxies. 
{\label{fig:lxsfr}}%
}
\end{center}
\end{figure*}

In Fig.\,\ref{fig:lxsfr} we show the \lx/SFR vs.\ sSFR for the 93 eFEDS-HECATE sources with SFR and stellar mass measurements. The symbols represent the classifications determined in Sect. \ref{sec:mclass}. The scaling relation from \citet{lehmer07-16} is plotted as a solid black line with dispersion in grey and represents the XRB+hot gas emission of local ($z\sim0$) normal galaxies (subsample from \citealt{lehmer11-10}) and stacked normal galaxy subsamples from the \chandra\ Deep Field South (CDF-S). 
The local ($z\sim0$) normal galaxy subsample from \citet{lehmer11-10} was for 2--10 keV XRB emission only, so \citet{lehmer07-16} applied a bandpass correction to obtain the 0.5--2 keV emission and added the hot gas contribution based on the results of \citet{mineo11-12}. \citet{mineo11-12} only studied late-type galaxies, although over a broad range of SFR from $\sim0.1-17$ \sfr, therefore this sample accurately reflects only the high-sSFR region of the scaling relation. The CDF-S stacked subsamples were derived from the observed-frame 0.5--1 keV emission, which probes the rest-frame 0.5--2 keV band emission, and thus includes the integrated hot gas + XRB emission. The \lx/SFR-sSFR scaling relation has generally been used to portray the connection between only the XRB emission of normal galaxies based on their SFR (scales with HMXB emission) and stellar mass (scales with LMXB emission).

Figure\,\ref{fig:lxsfr} shows that the star-forming galaxies have a large dispersion about the relation, particularly the dwarf galaxies at high-sSFR. However, the locus of star-forming galaxies is consistent with the scaling relation. Somewhat surprisingly, the AGN-classified galaxies are also mostly consistent with the scaling relation, with the exception of the lowest-sSFR galaxy (eFEDS ID 150) that is well above the relation. The AGN classifications are relevant in the broader context of normal galaxy properties as these are low-luminosity AGN (see Fig.\,\ref{fig:histlxssfrclass}), and yet appear to be indistinguishable from the locus of normal early-type galaxies at low-sSFR, scattering above and below the relation. There are also a few AGN and composites that overlap the lower-end of the high-sSFR region (defined in Sect. \ref{sec:mclass}), further motivating the need for robust spectroscopic classifications of galaxies from \ero\ surveys (e.g., eFEDS, eRASS). When plotting the \lx-SFR and \lx-\mstar\ relations we found similar trends showing an offset in both the dwarf galaxy population and a few of the most luminous AGN. We do not perform fitting of the \lx/SFR-sSFR relation for eFEDS-HECATE normal galaxies because we are effectively limited to a small region of high-sSFR normal galaxies.

\subsection{Scaling With Metallicity}\label{sec:scal2}

\begin{figure*}
\begin{center}
\begin{tabular}{cc}
\includegraphics[width=0.5\textwidth]{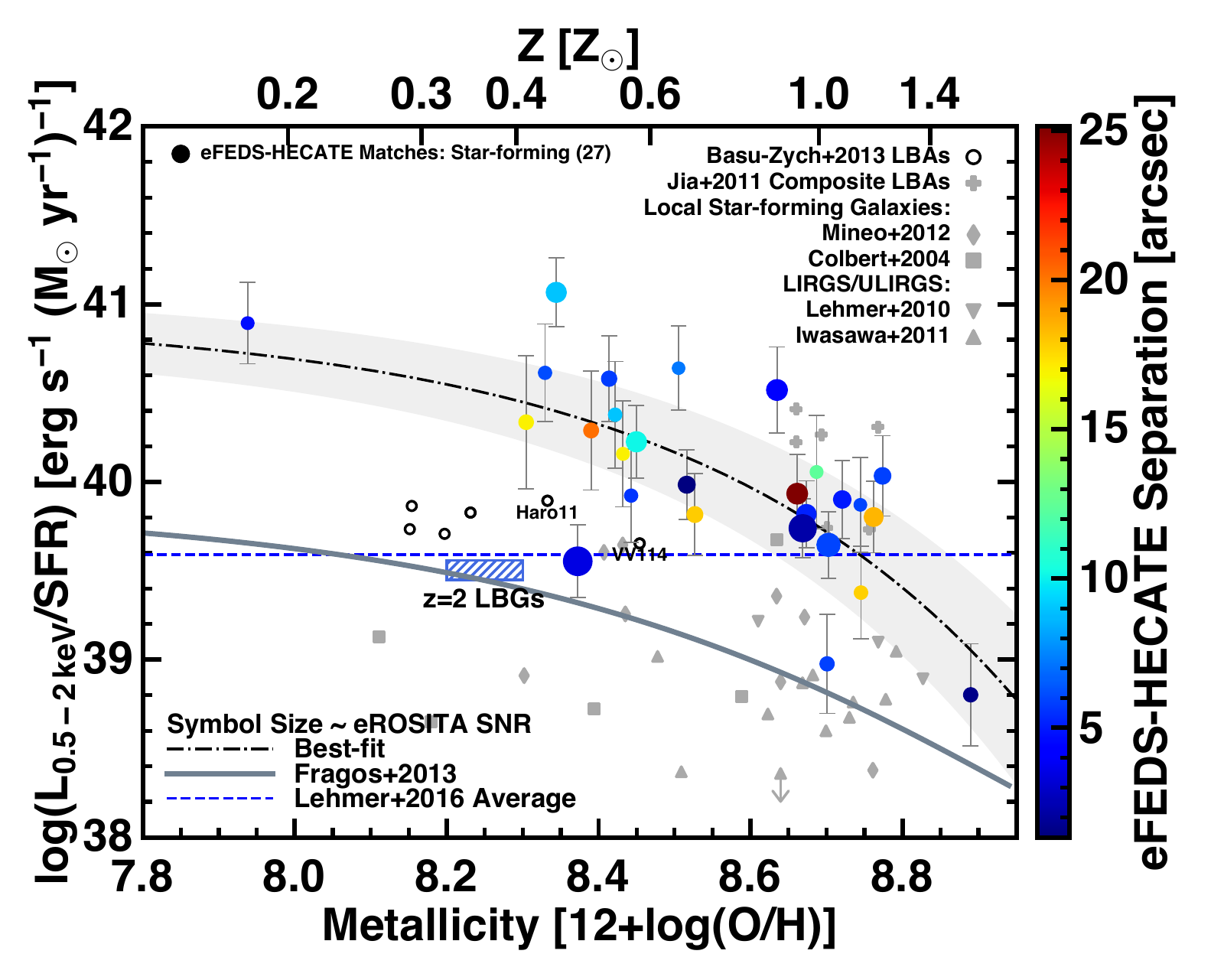}
\includegraphics[width=0.5\textwidth]{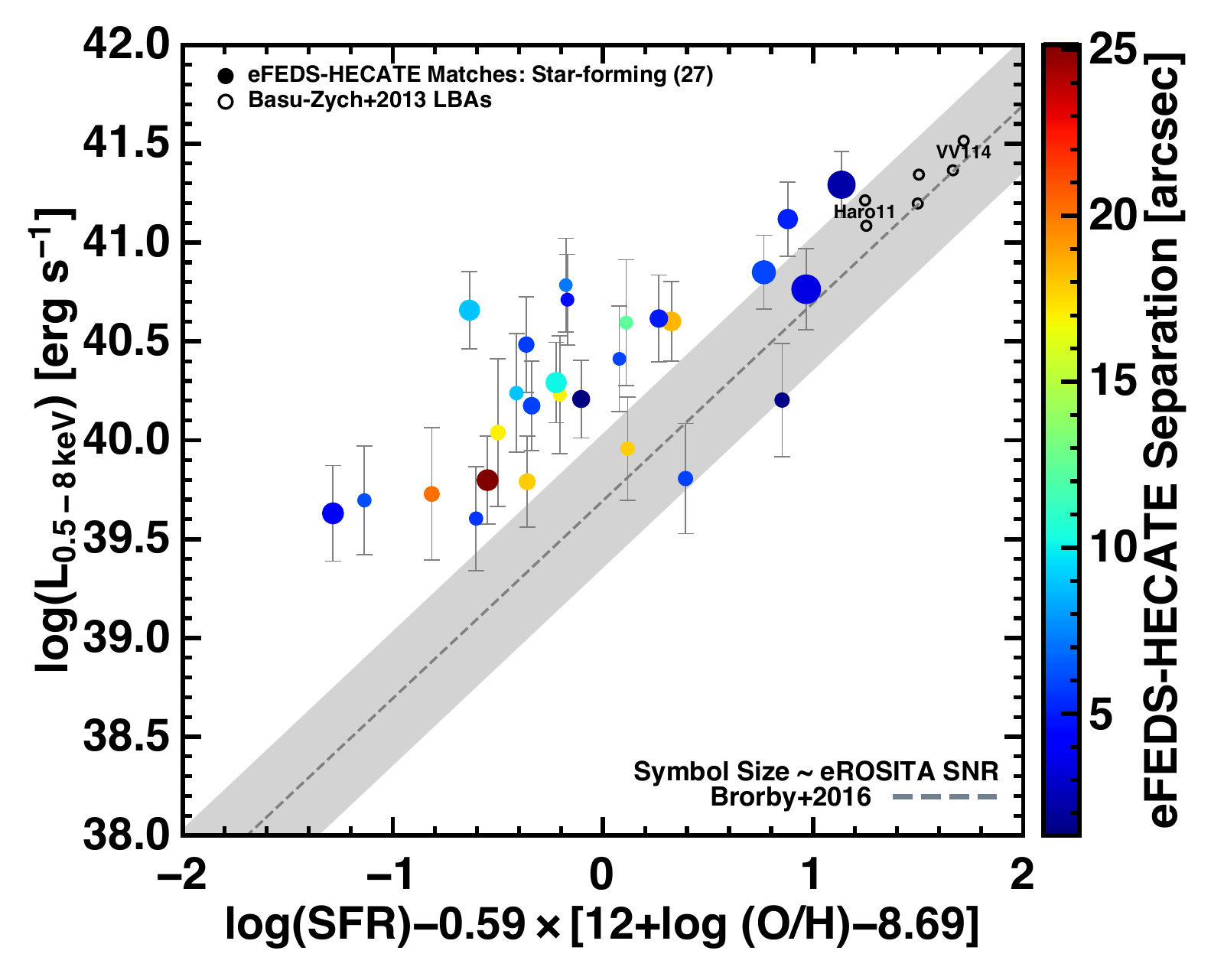}
\end{tabular}
\vspace{0.in}
\caption{\textit{Left:} \lx/SFR as a function of galaxy gas-phase metallicity for the eFEDS-HECATE galaxies classified as star-forming. \lx\ represents the integrated galaxy X-ray luminosity (XRBs + hot gas). We overlay Lyman break analogues from \citep[][unfilled black circles]{basu-zych09-13} and \citet[][filled grey plusses]{jia04-11}, local star-forming galaxies from \citet[][filled grey squares]{colbert02-04} and \citet[][filled grey diamonds]{mineo01-12}, and luminous/ultraluminous infrared galaxies from \citet[][inverted filled grey triangles]{lehmer11-10} and  \citet[][filled grey triangles]{iwasawa05-11}. The blue hatched region shows the range of values for $z=2$ Lyman break galaxies \citep{basu-zych09-13}. The theoretical prediction from \citet{fragos02-13} for HMXBs is shown as a solid dark grey line. The dashed blue line represents the average integrated (hot gas + HMXBs) $L_{0.5-2\rm{\ keV}}$/SFR of normal galaxies from \citet{lehmer07-16}. We do not have metallicity estimates for eFEDS-HECATE galaxies classified using 6dF spectra (Sect. \ref{sec:mclass}). The dot-dashed black line with grey dispersion shows the best-fit to the eFEDS-HECATE star-forming galaxies. The sample shows elevated \lx/SFR across metallicity, especially at low metallicity. This may be due to an obscured AGN, excessive  contribution from hot gas, or increased HMXB emission (see Sect. \ref{sec:scal2}). \textit{Right:} \lx-SFR-metallicity relation from \citet{brorby04-16} showing the best-fit from their paper for integrated $L_{0.5-8\rm{\ keV}}$ Lyman break analogues (dashed grey line with dispersion in dark grey). The offset between the eFEDS-HECATE galaxies and the relation for Lyman break analogues is pronounced towards low-SFR and low-metallicity. 
{\label{fig:metal}}%
}
\end{center}
\end{figure*}

In the left panel of Fig.\,\ref{fig:metal} we show the integrated (XRBs + hot gas) $L_{0.5-2\rm{\ keV}}$/SFR as a function of galaxy gas-phase metallicity $12+\log\rm{(O/H)}$ for \ero-detected star-forming galaxies. This metric is used to study the relation between HMXB emission (obtained from the 2--8 or 2--10 keV energy band) and metallicity in starburst galaxies \citep[e.g.][]{basu-zych09-13, douna07-15, brorby04-16, lehmer01-21}.  A metallicity dependence of \lx/SFR is expected due to HMXBs being more numerous and more luminous with decreasing metallicity, since weaker stellar winds allow more mass retention and tighter binary orbits, as demonstrated in X-ray binary population synthesis models \citep{linden12-10,fragos01-12}. We do not have metallicity estimates for eFEDS-HECATE galaxies classified using 6dF spectra (Sect. \ref{sec:mclass}), and therefore only \metng/\sfgal\ star-forming galaxies were studied here. As a result of only 2 normal galaxy detections in the \ero\ hard band (2--5 keV), we were unable to study HMXB-only emission from the population of normal galaxies in \ero. 
Thus data from previous surveys of Lyman break analogues, local star-forming galaxies, and luminous/ultraluminous infrared galaxies, that report HMXB-only emission, are not an accurate comparison to the eFEDS-HECATE sample. Nonetheless, we overplot galaxies from these various surveys to establish a reference sample. Bandpass corrections have been applied to convert from the 2--8 or 2--10 keV energy bands to 0.5--2 keV using the power-law spectral model defined at the end of Sect. \ref{sec:intro}. 	
We also included the 0.5--2 keV theoretical prediction from \citet{fragos02-13} for HMXBs (solid dark grey line) and a range of values from $z=2$ Lyman break galaxies \citep[blue hatched region][]{basu-zych09-13}. The dashed blue line represents the average integrated (hot gas + HMXBs) $L_{0.5-2\rm{\ keV}}$/SFR of normal galaxies from \citet{lehmer07-16}.

We fitted the eFEDS-HECATE galaxies from Fig.\,\ref{fig:metal} with the parametrization in equation \ref{eq:metalfit}. 
\begin{equation}
\log{\left(\frac{L_{\rm{X}}/\rm{SFR}}{\rm{erg\ s}^{-1}/M_{\odot} \rm{\ yr}^{-1}}\right)} = a\times(12+\log(\rm{O/H}))+b \label{eq:metalfit}
\end{equation}
Using the \textit{glm} function from the Python \texttt{statsmodels} package, we found $a=-1.19 \pm 0.18$ and $b=40.95\pm 0.15$ and dispersion about the fit of 0.28 dex. The dot-dashed black line with grey dispersion shows the best-fit to the eFEDS-HECATE star-forming galaxies. The eFEDS-HECATE sample exhibits elevated \lx/SFR across the metallicity range, especially at low metallicity. Comparing with the constant \lx/SFR integrated galaxy emission from \citet{lehmer07-16}, the cluster of eFEDS-HECATE galaxies near Z$_{\odot}$ are mostly within the dispersion of the relation (0.37 dex). However, the discrepancy is much more pronounced at lower metallicities.

What is the origin of the excess \lx/SFR seen in the eFEDS-HECATE sample, particularly at low metallicity? To attempt to account for this elevated emission, we subtracted estimates of the hot gas and LMXB emission from each eFEDS-HECATE star-forming galaxy based on the scaling relations from \citet{mineo11-12} and \citet{gilfanov03-04}, respectively. This allowed us to compare the estimate for HMXB-only emission from eFEDS-HECATE galaxies to the theoretical relation from \citet{fragos02-13} and data from previous surveys of Lyman break analogues, local star-forming galaxies, and luminous/ultraluminous infrared galaxies. This correction had a negligible effect on the total \lx/SFR of most sources, with the most extreme decrease being $\sim0.1$ dex. Therefore, unless the eFEDS-HECATE star-forming galaxies are part of a rare population with increased hot gas contribution that does not follow the known scaling relation, or have obscured AGN emission in the optical that is detected in the X-ray, these galaxies must have elevated HMXB emission due to low metallicity beyond what was already observed by previous studies of low-metallicity galaxies.

In the right panel of Fig.~\ref{fig:metal} we show the integrated (XRBs + hot gas) $\log(L_{0.5-8\rm{\ keV}})$ as a function of $\log(\rm{SFR})-0.59\times[12+\log(O/H)-8.69]$ for \ero-detected star-forming galaxies. This is the \lx-SFR-metallicity relation from \citet{brorby04-16}, which was fitted to the integrated emission of Lyman break analogue galaxies. We converted the \ero\ 0.5--2 keV flux to 0.5--8 keV to match the results from \citet{brorby04-16} using the same power-law model as above. 
This comparison is more accurate than the left panel of Fig.\,\ref{fig:metal} because it represents the observed integrated emission of galaxies at various metallicities that have been shown to exhibit elevated HMXB emission. 
The \ero\ star-forming galaxies with high SFRs are consistent with the \lx-SFR-Z relation whereas at low SFR they exhibit increased scatter and are elevated above the relation. This may be the result of an \ero\ selection effect where the rarest and brightest star-forming galaxies are detected. The galaxies showing significant scatter have lower stellar masses ($\lesssim10^{9}$ \msun) and SFRs ($<0.5$ \sfr), which resemble dwarf galaxies harbouring a single ULX that dominates the integrated X-ray emission. While these stochastic effects at low stellar masses and SFRs are known \cite{justham06-12, lehmer01-21}, the complete \ero\ survey will provide valuable measurements of the scatter at the most luminous end.

\section{Discussion}\label{sec:disc}

The results from studying normal galaxies in the eFEDS field are mixed when compared with predictions from simulations and scaling relations. SIXTE simulations from \citet{basu-zych08-20} were not consistent with observations, when comparing the overall number of normal galaxy detections and their morphological distribution. The findings also indicate some agreement with scaling relations but inconsistency at high-sSFR. The key results are (1) eFEDS observations found 1.9 times more normal galaxies than expected from simulations, and these were, except for one, all late-type galaxies, as opposed to an approximately even split in morphological type from simulations and (2) Star-forming galaxies nearer the dwarf mass regime at high-sSFR were found to have excess \lx/SFR relative to both sSFR and metallicity when compared with observational and theoretical expectations. In this section we discuss possible explanations for these results and their implications for studies of normal galaxies with \ero\ and beyond.

\subsection{Quantifying Normal Galaxies with \ero}\label{sec:dpred}

\begin{figure}%[!ht]

\begin{center}
\includegraphics[width=0.5\textwidth]{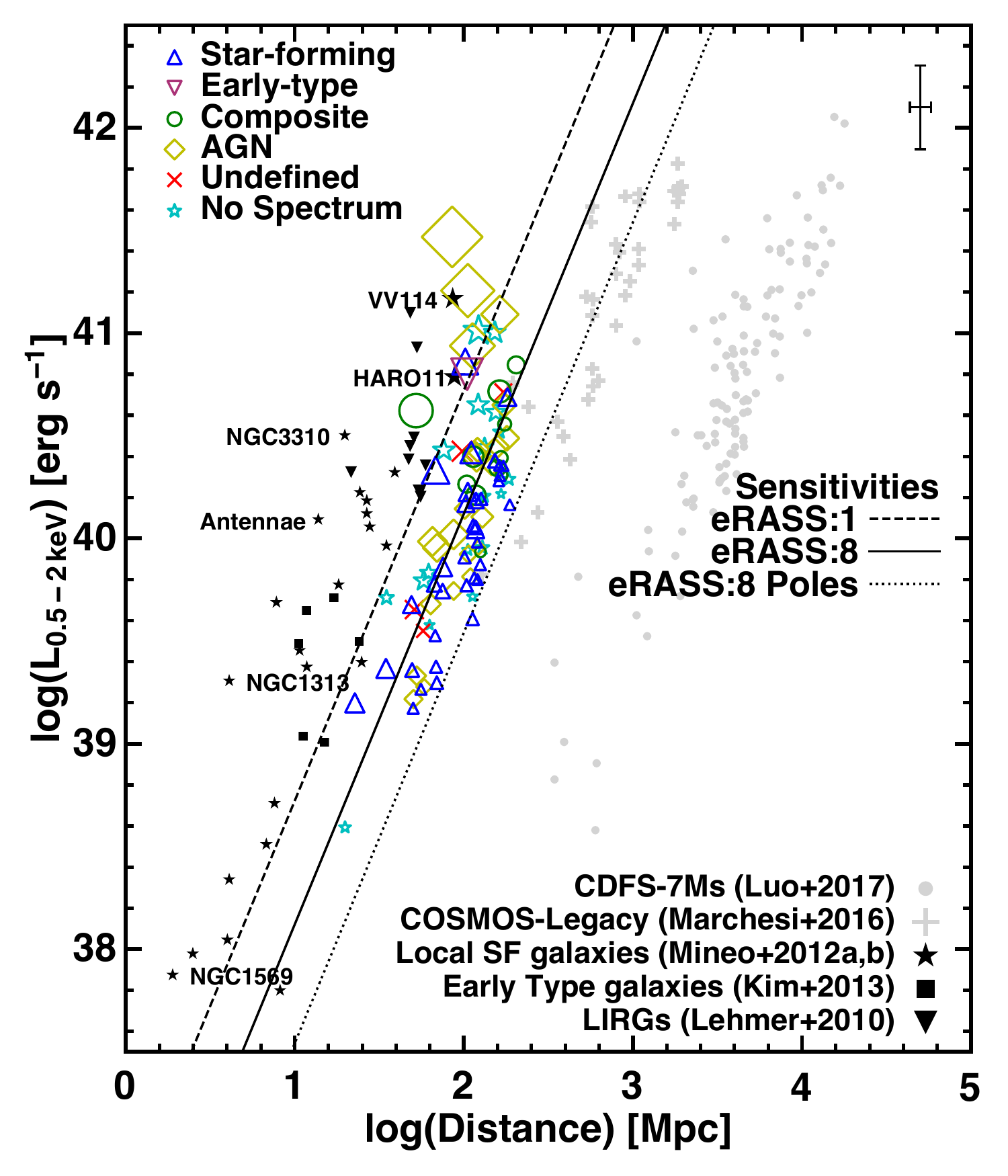}
\vspace{0.in}
\caption{\lx\ vs.\ Distance for eFEDS X-ray sources matched with HECATE galaxies and classified using optical spectra (see Sect. \ref{sec:mclass} for details). We show the \ero\ sensitivity limits for one epoch of the all-sky survey (eRASS:1, $\approx250$ s), the full 4-year survey (eRASS:8, $\approx2$ ks), and the full 4-year survey at the ecliptic poles ($\approx20$ ks). As expected, the eFEDS sources lie approximately at/above the eRASS:8 line, within uncertainty. We show nearby star-forming \citep[black filled stars;][]{mineo01-12, mineo11-12} and early-type \citep[black filled squares;][]{kim10-13} galaxies. A few well-known galaxies are labelled with their names for reference. Luminous Infrared galaxies \citep[LIRGS;][]{lehmer11-10} are shown as black filled inverted triangles. We also show galaxies from the COSMOS Legacy Survey \citep[grey plusses;][]{marchesi10-16} and CDF-S 7\,Ms survey \citep[filled grey circles;][]{luo01-17}. \ero\ probes a region of \lx$-D$ parameter space that has not been well studied. \label{fig:lxdist}}	
\end{center}
\end{figure}

To place the eFEDS sample in the context of the broader normal galaxy surveys, we show the eFEDS X-ray source luminosity as a function of distance to the matched host galaxy for eFEDS-HECATE matches in Fig. \ref{fig:lxdist}. The eFEDS galaxies are consistent with the sensitivity limit of eRASS:8, as expected, given the slightly improved flux limit attained by eFEDS of $\sim$\ten{7}{-15} \esc. Galaxies classified as AGN and composite compose many of the most luminous sources (see right panel of Fig. \ref{fig:histlxssfrclass}), whereas star-forming galaxies dominate the lower-end of the scale. When compared with nearby and high-redshift galaxies, the eFEDS sample (and subsequently eRASS) probes a region of \lx$-D$ parameter space that has not been studied previously due to limitations of current and past observatories. In particular, many of the high-sSFR dwarf galaxies detected thus far in eFEDS-HECATE would require targeted observations by existing X-ray telescopes, which would be expensive a priori based on the expected fluxes from the X-ray scaling relations. \ero\ will make significant advancements in the study of normal galaxies by increasing the sample size and detecting rare populations, such as metal-poor dwarf starbursts (green peas, blue compact dwarfs) and Wolf-Rayet galaxies.

While \normgal normal galaxies were identified from the \nmat eFEDS-HECATE matches, \ero\ observations of the eFEDS field also detected \nefeds X-ray sources, most of which were AGN. Planned optical spectroscopic follow-up of these X-ray sources will lead to the classification of additional normal galaxies, particularly star-forming galaxies, due to bias in the diagnostic criteria as explained in Sect. \ref{sec:mclass} and \ref{sec:charsixt}. The eFEDS-HECATE matches were detected out to $D\approx200$ Mpc, the approximate distance limit of the HECATE galaxy catalogue. The completeness of HECATE decreases with increasing distance, estimated at 50\% at $D\sim170$ Mpc in terms of $L_{\rm{B}}$ and \mstar. In terms of SFR, the completeness is much lower at 50\% for $30<D<150$ Mpc. There were 13 dwarf galaxies from HECATE (none matched to an eFEDS X-ray source) not plotted in Figure \ref{fig:sfrmass} due to their low masses (\mstar\ $\leq10^{7}$ \msun), which had sSFR distributed between .01 Gyr$^{-1}$ to 1 Gyr$^{-1}$ and $D<3$ Mpc. Past studies of X-ray emission from dwarf galaxies was mainly due to targeted observations, and therefore was limited and biased. Considering that the completeness of HECATE significantly affects low-mass (low-luminosity) galaxies, and approximately half of the normal galaxies detected in eFEDS were star-forming galaxies in the dwarf regime, this population is the most likely to dominate future identifications in the eFEDS field.

\subsection{Outliers in The Star-Forming Regime}\label{sec:outlie}

The comparison between detections of normal galaxies in SIXTE simulations \citep{basu-zych08-20} and eFEDS observations summarised in Table \ref{tab:sixtcomp} showed a significant discrepancy. eFEDS observations detected 1.9 times more normal galaxies than predicted from simulations, when scaled to the entire sky and calculated at identical flux and distance limits. This disagreement likely stems from the stochastic nature of low-mass galaxies near the dwarf regime (\mstar\ $\lesssim10^{9}$ \msun). Fig. \ref{fig:sfrmass} shows that almost half of the detected star-forming galaxies in eFEDS have stellar masses centred at $\sim10^{9}$ \msun. The scaling relations used to predict XRB and hot gas emission from normal galaxies have mostly excluded dwarf galaxies \citep[e.g.][\mstar\ $\lesssim 10^{9}$ \msun, SFR $\lesssim 0.1$ \sfr]{lehmer07-16}, mainly because they are less luminous, making detection difficult, and thus do not contribute substantially to the global X-ray emission of the Universe. However, a small subset of the dwarf galaxy population is vigorously star-forming with high-sSFR, showing elevated \lx/SFR (e.g., Fig. \ref{fig:lxsfr}). This offset was noticed by \citet{vulic09-18} when investigating XRB emission from galaxies with \nustar. The increased X-ray emission from some dwarf galaxies has been attributed to their low-metallicity, which increases the number and mass of compact objects and thus the total X-ray emission from XRBs \citep[e.g.][]{basu-zych01-13, basu-zych09-13, douna07-15, basu-zych02-16, brorby04-16, fornasini11-19, lehmer01-21}. The integrated X-ray emission from these low-metallicity dwarf galaxies is usually dominated by one source -- a ULX with \lx\ $>10^{39}$ \es\ \citep[e.g.][]{walton08-13, walton06-15}. Studies have shown that the numbers of ULXs increase with decreasing metallicity \citep{mapelli10-10, prestwich06-13, brorby07-14, douna07-15, basu-zych02-16, kovlakas11-20}. The stochastic nature of both low-metallicity dwarf galaxies in the local Universe and the formation of ULXs within these galaxies, combined with observational completeness issues, has made this population difficult to study to date. These characteristics, particularly the lack of empirical relations to predict the integrated X-ray emission from dwarf galaxies, resulted in part of the increased number of late-type galaxies detected in eFEDS observations. 
The flux-limited nature of eFEDS can be interpreted in this context as \ero\ having selected the brightest sources compared to the undetected population at similar metallicity, stellar mass, and SFR. While efforts are underway to measure the effects of stochasticity in the X-ray emission at low SFRs, stellar masses, and metallicities, the small numbers of such objects is a limiting factor \citep{kouroumpatzakis04-20, lehmer01-21}. By detecting larger samples of galaxies with these conditions, the rarest and most luminous low-metallicity dwarf galaxies in their respective metallicity, stellar mass, and SFR bins, \ero\ would significantly improve the current constraints on the scatter in scaling relations.

\subsection{Early-Type Galaxies: Missing or Misunderstood?}\label{sec:etype}

Further examination of the discrepancy between eFEDS observations and simulations reveals the paucity of early-type galaxy detections in eFEDS observations. Predictions of the integrated X-ray emission from early-type galaxies were dominated by the hot gas component, which carried large uncertainties from the known scaling relations. By extension, this incorporated large uncertainties on the total number of expected early-type galaxy detections, and may be a viable explanation for the lack of detections from eFEDS observations. 
However, a more likely explanation for the lack of early-type galaxy detections via eFEDS observations is that early-type galaxies were detected, except they hosted, and thus their X-ray emission was dominated by, low-luminosity AGN. The early-type region of Fig. \ref{fig:lxsfr} contains one confirmed normal early-type galaxy, 7 sources with no optical spectrum, 2 sources with low-quality spectra that prevented accurate classifications, 3 sources classified as composites, and 18 AGN. The early-types that were detected and classified were almost all AGN emitters (82\%). Studies of eRASS early-type galaxies must accurately assess their designation as normal or AGN-hosting to avoid misclassification.

Lastly, Fig. \ref{fig:lxdist} shows that nearby early-type galaxies from \citet[][filled black squares]{kim10-13} are located at the lowest luminosities probed by \ero\ in the eFEDS field, between $10^{39}$ and \ten{5}{39} \es. Even though late-type galaxies are detected in this \lx\ range, including a few unclassified sources, the early-type population may be underluminous in the eFEDS field, or preferentially host low-luminosity AGN. Thus it is not surprising that early-type normal galaxies, whether similar to this nearby sample or simply faint, are not detected. In addition to uncertainties on the hot gas component, studies of early-type galaxies have shown that the LMXB X-ray emission per unit stellar mass decreases with stellar age \citep{lehmer07-14, lehmer12-17}, and that LMXB emission per optical luminosity increases with globular cluster specific frequency \citep{humphrey12-08, boroson03-11}. While some subset of the eFEDS sample may harbour relatively old stellar populations and low globular cluster specific frequencies, contributing to a reduction in the total X-ray emission, the eFEDS field is large enough to avoid these selection effects and provide an unbiased, representative sample of early-type galaxies. Using optical/IR colour diagnostics, Salvato et al. (2021, submitted) identified a population of brightest cluster galaxies that were consistent with S0 and elliptical galaxy spectral energy distribution template tracks. Whether these sources are true passive (normal) galaxies or AGN was not evident and required further investigation.

\subsection{Implications for eRASS Normal Galaxy Detections}\label{sec:implic}

The first results from \ero\ show that there is still much to be learned about normal galaxies in the Universe. The predictions from SIXTE simulations based on known scaling relations of XRBs and hot gas in normal galaxies were inaccurate in predicting exactly \textit{which} galaxies would be detected (only 3 galaxies were detected by \ero\ observations and SIXTE simulations), as well as the overall number of detections. A more comprehensive analysis of the hot gas emission from normal galaxies is required to understand the integrated X-ray emission from normal galaxies and model the X-ray output across cosmic time. 
In addition, the lack of early-type normal galaxy detections from eFEDS observations was unexpected based on simulations from \citet{basu-zych08-20}, although optical spectroscopic classifications indicate that most host low-luminosity AGN. Understanding the X-ray emission from early-type galaxies is critical to evaluate the results of eRASS surveys and discerning why most within eFEDS are AGN-hosting and not normal galaxies. 
The high rate of AGN identifications among early-types represents a tracer of the SMBH accretion rate in this population. Even within the \fov deg$^{2}$ eFEDS field, we can conclude that the SMBH accretion rate in early-type galaxies must be large, even if at low-luminosities of \lx\ $\lesssim$ \ten{3}{41} \es\ when compared to luminous AGN. 

The predictions also did not foresee the number of low-mass (dwarf) galaxy detections that were found within eFEDS, partly because we do not understand the ULX-hosting rate and metallicity effects of this population \citep[cf.][]{kovlakas11-20}. The scaling relations for XRBs and hot gas that are used for predicting X-ray emission from galaxies do not extend into the dwarf regime, hampering our ability to predict the properties of these galaxies. Improving scaling relations for the integrated X-ray emission of normal galaxies, particularly the hot gas contribution and effects of stellar age and metallicity, will help in understanding the total X-ray output of normal galaxies in the local Universe. The elevated values of \lx/SFR, when compared with empirical relations and theoretical predictions, (Figs. \ref{fig:lxsfr} and \ref{fig:metal}) for dwarf galaxies are likely the result of increased HMXB/ULX emission due to low-metallicity. However, the presence of an intermediate mass BH in these dwarf starbursts cannot be ruled out based on their identification in previous studies \citep[e.g.][]{baldassare08-15, mezcua-17, huang04-19}. 
This elevated X-ray emission from dwarf galaxies could be attributed to active massive ($10^{4.1-5.8}$ \msun) BH that may be off-nuclear \citep{reines01-20}. The overall implications for eRASS normal galaxies are that many detections will be late-type star-forming galaxies, and early-type normal galaxies will be rarer than predicted due to the presence of low-luminosity AGN.

\section{Conclusions}\label{sec:conc}

Based on its expected 4-year survey coverage and sensitivity, \ero~will significantly increase the number of X-ray detected normal galaxies and further our understanding of the X-ray emission produced in galaxies. In this paper, we report on the detection of \normgal normal galaxies in the \ero\ eFEDS survey, representing the first view of X-ray emission from XRBs and the hot ISM at the full depth of the eRASS survey that is being conducted over the interval 2019--2023. We summarise our main results as follows: 
\begin{itemize}
    \item We found \nmat eFEDS X-ray sources that were matched within the \dtf\ ellipse of galaxies from the HECATE catalogue. These associations are consistent with the counterpart identifications determined by Salvato et al. (2021, submitted) in all but 7 cases, for detection likelihoods $\geq$6. Using optical spectroscopy from SDSS and 6dF, we classified sources as galaxies (late-type, early-type), composite, and AGN. We identified \normgal normal galaxies, comprised of \sfgal late-type (star-forming) galaxies and \egal early-type galaxy. 
    \item Contrary to expectations from the simulations of \citet{basu-zych08-20}, we found 1.9 times more normal galaxies, dominated by an increased number of late-type galaxies and a dearth of early-type galaxies in the eFEDS X-ray detected sample, compared with those predictions. It was noted by \citet{basu-zych08-20} that the properties of the hot interstellar medium in ``less extreme"  (\eg $-11<\log \mathrm{sSFR}<-10$) galaxies is not well understood. Future research should focus on a careful investigation of early-type galaxies in eRASS, which were mostly found to host AGN in the eFEDS field. 
    \item Based on extrapolation from the eFEDS field, we expect at least 12\,500 normal galaxies will be detected in eRASS:8, dominated by star-forming galaxies. The absence of all but one early-type galaxy may be due to a high SMBH accretion rate at low luminosity, or differences in the hot gas prescriptions for this population, which was expected to dominate the X-ray emission of early-type galaxies in simulations. 
    \item We find general consistency between the \ero-detected sample and previous X-ray surveys in terms of correlations such as \lx-sSFR and \lx-metallicity, but do find large scatter, towards higher X-ray luminosities, at the high-sSFR end. This subsample represents dwarf galaxies that are low-metallicity and are generally dominated by one (or possibly a few within the \ero\ PSF) luminous source(s) (e.g., ULX, IMBH). As \ero\ will be populating the extreme end of such relations with more rare populations such as vigorous starbursts, more detailed studies of these rare galaxies will be possible. 
   
\end{itemize}
The follow-up of eFEDS X-ray sources is underway to confirm their nature and characterize their multiwavelength properties. eRASS, the first X-ray all-sky survey since ROSAT, will revolutionize our understanding of normal galaxies in the nearby Universe, detecting a statistically significant population free from the selection effect of targeted observations studying specific galaxies. 
Complementing deeper pencil-beam surveys (\eg the \chandra\ Deep Fields) whose power was in detecting distant and faint sources, \ero\ will significantly boost the detection numbers and survey completeness for the brightest, yet rarest, nearby galaxy populations which may be those most representative of the first, primordial, galaxies. Therefore, these studies will have far-reaching implications on galaxy formation and evolution, particularly the constraints placed on local analogues of high-redshift galaxies thought to contribute to the heating of the intergalactic medium.

\begin{acknowledgements}
A.R.B. acknowledges support from NASA under award number 80GSFC21M0002. 
This work is based on data from \ero, the soft X-ray instrument aboard SRG, a joint Russian-German science mission supported by the Russian Space Agency (Roskosmos), in the interests of the Russian Academy of Sciences represented by its Space Research Institute (IKI), and the Deutsches Zentrum f\"ur Luft- und Raumfahrt (DLR). The SRG spacecraft was built by Lavochkin Association (NPOL) and its subcontractors, and is operated by NPOL with support from the Max Planck Institute for Extraterrestrial Physics (MPE).

The development and construction of the \ero\ X-ray instrument was led by MPE, with contributions from the Dr.~Karl Remeis Observatory Bamberg \& ECAP (FAU Erlangen-Nuernberg), the University of Hamburg Observatory, the Leibniz Institute for Astrophysics Potsdam (AIP), and the Institute for Astronomy and Astrophysics of the University of T\"{u}bingen, with the support of DLR and the Max Planck Society. The Argelander Institute for Astronomy of the University of Bonn and the Ludwig Maximilians Universit\"{a}t Munich also participated in the science preparation for \ero.

Funding for the Sloan Digital Sky Survey (SDSS) has been provided by the Alfred P. Sloan Foundation, the Participating Institutions, the National Aeronautics and Space Administration, the National Science Foundation, the US Department of Energy, the Japanese Monbukagakusho, and the Max Planck Society. The SDSS Web site is http://www.sdss.org/. The SDSS is managed by the Astrophysical Research Consortium (ARC) for the Participating Institutions. The Participating Institutions are The University of Chicago, Fermilab, the Institute for Advanced Study, the Japan Participation Group, The Johns Hopkins University, Los Alamos National Laboratory, the Max-Planck-Institute for Astronomy (MPIA), the Max-Planck-Institute for Astrophysics (MPA), New Mexico State University, University of Pittsburgh, Princeton University, the United States Naval Observatory, and the University of Washington.

This research has made use of data obtained from the \chandra\ Source Catalog, provided by the \chandra\ X-ray Center (CXC) as part of the \chandra\ Data Archive. This research has made use of data obtained from the 4XMM \xmmn\ serendipitous source catalogue compiled by the 10 institutes of the \xmmn\ Survey Science Centre selected by ESA. This publication makes use of data products from the Wide-field Infrared Survey Explorer, which is a joint project of the University of California, Los Angeles, and the Jet Propulsion Laboratory/California Institute of Technology, funded by the National Aeronautics and Space Administration. This publication makes use of data products from the Two Micron All Sky Survey, which is a joint project of the University of Massachusetts and the Infrared Processing and Analysis Center/California Institute of Technology, funded by the National Aeronautics and Space Administration and the National Science Foundation. This research has made use of the NASA/IPAC Extragalactic Database (NED) which is operated by the Jet Propulsion Laboratory, California Institute of Technology, under contract with the National Aeronautics and Space Administration. We acknowledge the usage of the HyperLeda database (\url{http://leda.univ-lyon1.fr}).

\end{acknowledgements}

\textit{Facilities: } \srg\ (\ero)

\textit{Software: } astropy \citep{astropy-collaboration10-13, the-astropy-collaboration09-18}, TOPCAT \citet{taylor12-05}

\longtab{
{\fontsize{7.5}{9}\selectfont	%	use 1.2 times the font size for the array stretch spacing between lines

{\setlength\tabcolsep{3pt}
\begin{landscape}
\begin{longtable}{@{}llrrrrrrrrrlrrcrrrrr@{}}
\caption{Source Characteristics for eFEDS-HECATE Matches\label{tab:srcchar}}	\\
\hline\hline
PGC & ObjectName & HECATE RA & HECATE Dec & PA & S$_{\rm{maj}}$ & S$_{\rm{min}}$ & $D$ & $D_{\rm{err}}$ & SFR & \mstar & eFEDS ID & eFEDS RA & eFEDS Dec & Catalogue & Flux & $\sigma$ & Sep & Class & $12 + \log(\mathrm{O/H})$ \\
	&	&	\multicolumn{2}{c}{(J2000.0)}	&	(\degr)	&	\multicolumn{2}{c}{(\arcmin)}	&	\multicolumn{2}{c}{(Mpc)}	&	(\sfr)	&	(10$^{9}$  \msun)	&	&	\multicolumn{2}{c}{(J2000.0)}	&	&	\multicolumn{2}{c}{($10^{-15}$ \esc)}	&	(\arcsec)	&	&	\\
\cmidrule(lr){3-4} \cmidrule(lr){6-7} \cmidrule(lr){8-9} \cmidrule(lr){13-14} \cmidrule(lr){16-17}
(1) &	(2) & (3)	&	(4)	&	(5)	&	(6) & (7) & (8)	&	(9)	&	(10)	&	(11)	&	(12)	&	(13)	&	(14)	&	(15)	&	(16)	&	(17)	&	(18)	&	(19)	&	(20)	\\	
\hline
\endfirsthead
\caption{continued.}\\
\hline\hline
PGC & ObjectName & HECATE RA & HECATE Dec & PA & S$_{\rm{maj}}$ & S$_{\rm{min}}$ & $D$ & $D_{\rm{err}}$ & SFR & \mstar & eFEDS ID & eFEDS RA & eFEDS Dec & Catalogue & Flux & $\sigma$ & Sep & Class & $12 + \log(\mathrm{O/H})$ \\
\hline
\endhead
\hline
\endfoot
23721 & PGC023721 & 126.8825610 & -1.6252583 & 89 & 0.271 & 0.204 & 152.6 & 15.5 & 4.2527 & 56.77184 & 5842 & 126.8822975 & -1.6245205 & M & 36.145 & 8.516 & 2.8 & -99 & -99.00 \\
23749 & UGC04430 & 127.0332750 & -1.5386406 & 141 & 0.516 & 0.258 & 105.7 & 23.4 & 1.5948 & 113.13803 & 27927 & 127.0355509 & -1.5380081 & M & 6.579 & 3.104 & 8.5 & -99 & -99.00 \\
1119085 & PGC1119085 & 127.5480240 & -1.4188387 & 150 & 0.204 & 0.138 & 158.1 & 17.2 & 3.7762 & 34.70771 & 7851 & 127.5485521 & -1.4184034 & M & 7.371 & 2.520 & 2.5 & 3 & -99.00 \\
23889 & PGC023889 & 127.7426205 & -0.0926050 & 83 & 0.229 & 0.153 & 152.7 & 15.6 & 4.4903 & 55.16358 & 24621 & 127.7408039 & -0.0921771 & M & 8.568 & 3.079 & 6.7 & 0 & -99.00 \\
23954 & PGC023954 & 128.1233685 & -0.9030143 & 0 & 0.290 & 0.251 & 125.2 & 16.0 & 8.6088 & 59.46416 & 20749 & 128.1230299 & -0.9047055 & M & 3.982 & 2.210 & 6.2 & 0 & -99.00 \\
152776 & PGC152776 & 128.1951090 & -0.3296365 & 148 & 0.283 & 0.148 & 165.8 & 20.2 & 3.0422 & 29.80169 & 23962 & 128.1938937 & -0.3276043 & M & 7.508 & 2.779 & 8.5 & 3 & -99.00 \\
23973 & UGC04467 & 128.1952500 & 0.2281389 & 164 & 0.598 & 0.273 & 132.7 & 15.9 & 4.4275 & 185.21416 & 12742 & 128.1973316 & 0.2276456 & M & 13.238 & 3.706 & 7.7 & -99 & -99.00 \\
1281998 & PGC1281998 & 128.5955775 & 5.3481025 & 79 & 0.170 & 0.116 & 187.1 & 20.6 & 1.2845 & 11.40250 & 32440 & 128.5987396 & 5.3467623 & S & 3.487 & 2.278 & 12.3 & 0 & 8.69 \\
24068 & PGC024068 & 128.6321535 & 1.6661131 & 77 & 0.396 & 0.153 & 58.1 & 8.6 & 2.1717 & 11.26701 & 6790 & 128.6341833 & 1.6671734 & M & 15.424 & 4.001 & 8.2 & -99 & -99.00 \\
3996672 & SDSSJ083437.19+042816.2 & 128.6550135 & 4.4712029 & 90 & 0.203 & 0.148 & 55.8 & 8.6 & 0.0449 & 0.36040 & 30488 & 128.6557703 & 4.4696749 & S & 4.942 & 2.435 & 6.1 & 0 & 8.33 \\
1108322 & PGC1108322 & 128.7738810 & -1.8505761 & 60 & 0.248 & 0.113 & 184.5 & 49.3 & 0.5697 & 82.11586 & 4194 & 128.7718665 & -1.8502428 & M & 4.777 & 1.993 & 7.3 & -99 & -99.00 \\
24129 & NGC2616 & 128.8919310 & -1.8501282 & 146 & 0.713 & 0.491 & 104.7 & 22.7 & 1.2285 & 257.75492 & 1284 & 128.8915820 & -1.8498667 & M & 48.094 & 5.364 & 1.6 & 6 & -99.00 \\
180931 & PGC180931 & 128.9099775 & -1.2352970 & 65 & 0.311 & 0.118 & 179.8 & 16.6 & 8.3407 & 88.02133 & 13602 & 128.9084154 & -1.2354569 & M & 7.958 & 2.794 & 5.7 & 1 & -99.00 \\
3092142 & PGC3092142 & 128.9115690 & 4.7570398 & 77 & 0.310 & 0.277 & 58.3 & 8.5 & 0.1896 & 14.03225 & 22501 & 128.9110970 & 4.7580027 & M & 4.657 & 2.219 & 3.9 & 2 & 8.68 \\
24135 & PGC024135 & 128.9239800 & -1.7567672 & 141 & 0.427 & 0.331 & 128.8 & 14.9 & 2.3157 & 72.83403 & 21546 & 128.9254648 & -1.7582419 & M & 4.543 & 1.906 & 7.5 & -99 & -99.00 \\
24145 & PGC024145 & 128.9463750 & -1.9438366 & 53 & 0.275 & 0.132 & 154.7 & 15.3 & 0.7865 & 131.93583 & 5741 & 128.9473457 & -1.9444631 & M & 14.381 & 3.174 & 4.2 & -99 & -99.00 \\
24152 & UGC04491 & 128.9521320 & 1.7217086 & 57 & 1.052 & 0.367 & 61.9 & 11.4 & 0.9903 & 83.94757 & 13202 & 128.9518019 & 1.7242006 & M & 14.817 & 3.902 & 9.0 & -99 & -99.00 \\
24156 & NGC2618 & 128.9731530 & 0.7071007 & 141 & 1.035 & 0.442 & 62.8 & 11.6 & -99.0000 & 157.18423 & 28041 & 128.9744703 & 0.7095372 & S & 8.030 & 3.146 & 10.0 & -99 & -99.00 \\
1262541 & PGC1262541 & 129.1285455 & 4.0376769 & 168 & 0.413 & 0.232 & 121.5 & 16.5 & 1.1729 & 76.13361 & 32636 & 129.1280188 & 4.0365659 & S & 5.100 & 2.366 & 4.4 & 1 & 8.70 \\
24198 & PGC024198 & 129.1622385 & -1.5317004 & 69 & 0.301 & 0.148 & 128.8 & 14.9 & 2.5214 & 62.63190 & 17053 & 129.1610011 & -1.5308608 & M & 6.426 & 2.497 & 5.4 & 1 & -99.00 \\
1099872 & PGC1099872 & 129.3014055 & -2.1679147 & 64 & 0.245 & 0.117 & 162.8 & 19.6 & 2.2760 & 98.52717 & 7833 & 129.3014073 & -2.1687107 & M & 16.409 & 4.239 & 2.9 & 3 & -99.00 \\
1208190 & PGC1208190 & 129.3482550 & 1.8575081 & 62 & 0.211 & 0.119 & 131.5 & 16.0 & 0.6721 & 1.19399 & 22740 & 129.3498070 & 1.8579402 & M & 7.768 & 2.901 & 5.8 & -99 & -99.00 \\
1108649 & PGC1108649 & 129.5290500 & -1.8385053 & 169 & 0.388 & 0.089 & 158.5 & 17.0 & 3.8545 & 44.42228 & 17437 & 129.5282261 & -1.8379313 & M & 9.348 & 3.052 & 3.6 & 1 & -99.00 \\
24317 & PGC024317 & 129.7182090 & 1.3991497 & 126 & 0.224 & 0.187 & 121.6 & 16.5 & 15.2372 & 39.18775 & 7303 & 129.7188481 & 1.3994767 & M & 25.290 & 4.873 & 2.6 & -99 & -99.00 \\
2816619 & PGC2816619 & 129.9133305 & 3.8284863 & 25 & 0.229 & 0.151 & 110.6 & 14.6 & 5.9332 & 21.04204 & 9121 & 129.9116806 & 3.8283433 & M & 17.923 & 4.164 & 5.9 & 0 & 8.70 \\
3450700 & SDSSJ084117.92+035306.8 & 130.3246575 & 3.8852359 & 0 & 0.176 & 0.166 & 115.3 & 15.7 & 0.2691 & 0.84520 & 20610 & 130.3224275 & 3.8863303 & M & 4.042 & 2.356 & 8.9 & 0 & 8.42 \\
24425 & NGC2644 & 130.3828155 & 4.9803368 & 12 & 0.822 & 0.324 & 19.9 & 4.9 & 0.3614 & 4.61954 & 25340 & 130.3823353 & 4.9817615 & M & 8.299 & 3.220 & 5.4 & -99 & -99.00 \\
24441 & PGC024441 & 130.4986200 & 0.9278011 & 0 & 0.208 & 0.129 & 163.2 & 19.2 & 1.7321 & 107.23632 & 22013 & 130.5006805 & 0.9299180 & M & 10.377 & 3.558 & 10.6 & -99 & 8.19 \\
153158 & PGC153158 & 130.5076890 & -1.7537775 & 118 & 0.220 & 0.157 & 163.5 & 20.0 & 10.7065 & 58.93761 & 27601 & 130.5073015 & -1.7510199 & S & 6.318 & 2.881 & 10.0 & 0 & -99.00 \\
24461 & PGC024461 & 130.6671645 & -0.5632081 & 21 & 0.356 & 0.255 & 126.7 & 15.6 & 2.0467 & 60.72774 & 6828 & 130.6640816 & -0.5636169 & M & 8.127 & 3.257 & 11.2 & 0 & -99.00 \\
213516 & PGC213516 & 130.8263085 & 3.6003989 & 81 & 0.197 & 0.122 & 112.8 & 14.6 & 0.1872 & 0.88269 & 29989 & 130.8216840 & 3.6012418 & M & 2.665 & 2.086 & 16.9 & 0 & 8.30 \\
24499 & UGC04553 & 130.8588555 & 3.6146130 & 31 & 0.489 & 0.208 & 112.8 & 14.6 & 6.3209 & 75.15505 & 19929 & 130.8562835 & 3.6171958 & M & 3.425 & 2.316 & 13.1 & -99 & -99.00 \\
4532355 & SDSSJ084643.28+023221.2 & 131.6802960 & 2.5391898 & 69 & 0.336 & 0.216 & 120.5 & 17.1 & 0.4414 & 13.81124 & 11246 & 131.6815433 & 2.5380812 & M & 15.580 & 4.178 & 6.0 & 2 & 8.68 \\
24658 & IC0521 & 131.6831685 & 2.5374212 & 79 & 0.466 & 0.335 & 117.7 & 15.7 & 1.9144 & 188.05646 & 11246 & 131.6815433 & 2.5380812 & M & 15.580 & 4.178 & 6.3 & 2 & 8.68 \\
24668 & PGC024668 & 131.7251700 & 2.5014672 & 117 & 0.194 & 0.156 & 118.1 & 16.0 & 1.9231 & 12.46751 & 15267 & 131.7238194 & 2.5013295 & M & 9.185 & 3.207 & 4.9 & 0 & 8.72 \\
1192549 & PGC1192549 & 131.8087170 & 1.3624002 & 60 & 0.310 & 0.242 & 173.8 & 34.3 & 2.5959 & 190.16492 & 9129 & 131.8083700 & 1.3634400 & M & 12.353 & 3.907 & 3.9 & 1 & 8.74 \\
24693 & PGC024693 & 131.8811295 & 2.6902814 & 154 & 0.406 & 0.144 & 120.1 & 17.1 & 1.2889 & 21.85500 & 29104 & 131.8825943 & 2.6895172 & S & 5.547 & 2.665 & 5.9 & 0 & 8.74 \\
24827 & PGC024827 & 132.5515920 & 0.6568154 & 145 & 0.245 & 0.163 & 174.5 & 18.2 & 2.0331 & 58.11490 & 21972 & 132.5505381 & 0.6546283 & M & 9.898 & 3.695 & 8.7 & 3 & 8.69 \\
24830 & UGC04625 & 132.5738310 & 3.4975092 & 0 & 0.839 & 0.839 & 122.2 & 17.7 & 3.0677 & 59.39215 & 1626 & 132.5749694 & 3.5107355 & M & 57.404 & 7.832 & 47.8 & -99 & -99.00 \\
24893 & UGC04640 & 132.9325530 & -2.1340313 & 6 & 1.291 & 0.440 & 49.1 & 7.1 & 0.6571 & 25.14708 & 13598 & 132.9284047 & -2.1306827 & M & 16.517 & 4.578 & 19.2 & 0 & -99.00 \\
3120822 & PGC3120822 & 133.2898110 & 2.2464099 & 12 & 0.283 & 0.153 & 146.1 & 16.6 & 0.6151 & 64.40996 & 15499 & 133.2906298 & 2.2470038 & M & 8.882 & 3.637 & 3.6 & 1 & 8.53 \\
24969 & PGC024969 & 133.4197080 & 1.5634810 & 85 & 0.279 & 0.211 & 167.4 & 20.6 & 3.8234 & 66.60377 & 19996 & 133.4213870 & 1.5612329 & M & 6.067 & 3.019 & 10.1 & 3 & 8.62 \\
1187281 & PGC1187281 & 133.5246270 & 1.1918714 & 65 & 0.183 & 0.128 & 181.3 & 17.7 & 7.4121 & 33.87165 & 16894 & 133.5233042 & 1.1923879 & M & 12.400 & 3.644 & 5.1 & 0 & 8.67 \\
25070 & UGC04672 & 133.9299990 & -2.2888718 & 0 & 0.411 & 0.411 & 162.0 & 18.7 & 12.6819 & 162.42692 & 30767 & 133.9274063 & -2.2903238 & S & 7.230 & 3.284 & 10.7 & 0 & -99.00 \\
25102 & NGC2706 & 134.0512065 & -2.5634743 & 168 & 0.836 & 0.283 & 34.9 & 9.0 & 4.4145 & 32.17867 & 22658 & 134.0496155 & -2.5630630 & M & 35.217 & 10.678 & 5.9 & -99 & -99.00 \\
25128 & UGC04684 & 134.1694950 & 0.3750087 & 157 & 0.644 & 0.512 & 34.6 & 6.3 & 0.2715 & 2.59048 & 7551 & 134.1754928 & 0.3785893 & M & 16.223 & 4.165 & 25.1 & 0 & 8.66 \\
25161 & NGC2713 & 134.3354310 & 2.9213206 & 106 & 1.683 & 0.573 & 65.3 & 11.5 & 4.0876 & 343.02725 & 8956 & 134.3346540 & 2.9227582 & M & 18.939 & 4.479 & 5.9 & 1 & 8.67 \\
1145379 & PGC1145379 & 134.4385980 & -0.3566063 & 49 & 0.196 & 0.071 & 118.2 & 16.0 & 0.2968 & 1.28529 & 20630 & 134.4373197 & -0.3575533 & M & 6.765 & 2.810 & 5.7 & 0 & 8.41 \\
25179 & PGC025179 & 134.4471600 & -0.1999482 & 0 & 0.286 & 0.267 & 117.9 & 15.8 & 3.3132 & 48.48312 & 10428 & 134.4463291 & -0.1998656 & M & 9.761 & 3.140 & 3.0 & 3 & 8.67 \\
25206 & PGC025206 & 134.6191515 & 0.0234503 & 152 & 0.323 & 0.222 & 119.1 & 16.8 & 3.1546 & 46.52990 & 5836 & 134.6197519 & 0.0233387 & M & 14.485 & 3.893 & 2.2 & 1 & 8.55 \\
4009917 & SDSSJ085907.12+031704.0 & 134.7797430 & 3.2844664 & 173 & 0.277 & 0.150 & 51.0 & 7.7 & 0.0873 & 0.16032 & 11897 & 134.7820665 & 3.2852303 & M & 14.461 & 3.810 & 8.8 & 5 & 8.34 \\
1147505 & PGC1147505 & 134.7900300 & -0.2721533 & 19 & 0.156 & 0.082 & 171.2 & 19.5 & 0.5179 & 1.79473 & 16467 & 134.7893183 & -0.2703052 & S & 6.445 & 2.764 & 7.1 & 0 & 8.51 \\
25259 & PGC025259 & 134.9543445 & 5.0608578 & 173 & 0.393 & 0.206 & 52.2 & 7.9 & 0.5973 & 3.33183 & 1034 & 134.9544330 & 5.0610130 & M & 128.500 & 14.019 & 0.6 & 3 & 8.66 \\
25363 & PGC025363 & 135.4382250 & 0.5218678 & 134 & 0.391 & 0.140 & 75.3 & 10.7 & 0.4319 & 5.25593 & 13787 & 135.4389846 & 0.5245683 & M & 10.705 & 2.948 & 10.1 & 0 & 8.45 \\
25391 & PGC025391 & 135.6224100 & 3.3849801 & 66 & 0.408 & 0.171 & 112.5 & 14.6 & 0.8105 & 110.03922 & 1399 & 135.6221996 & 3.3849392 & M & 57.235 & 6.993 & 0.8 & 2 & 8.65 \\
25436 & PGC025436 & 135.8890320 & 3.3758393 & 115 & 0.442 & 0.143 & 109.5 & 14.0 & 0.9698 & 67.17637 & 31517 & 135.8887725 & 3.3756358 & M & 4.541 & 2.550 & 1.2 & 1 & 8.55 \\
25444 & PGC025444 & 135.9038940 & -1.1301910 & 174 & 0.341 & 0.152 & 166.2 & 20.2 & 6.3117 & 62.71780 & 11089 & 135.9044896 & -1.1297054 & S & 4.989 & 2.357 & 2.8 & -99 & -99.00 \\
3354774 & SDSSJ090531.07+033530.3 & 136.3793985 & 3.5917907 & 55 & 0.147 & 0.079 & 161.1 & 18.1 & 0.2438 & 0.79960 & 6034 & 136.3783873 & 3.5909791 & S & 6.136 & 2.491 & 4.7 & 0 & 7.94 \\
25560 & PGC025560 & 136.6647930 & -0.8652896 & 88 & 0.244 & 0.143 & 125.4 & 15.8 & 4.2568 & 36.43716 & 30845 & 136.6678685 & -0.8662689 & S & 4.571 & 2.263 & 11.6 & 3 & -99.00 \\
25646 & NGC2765 & 136.9026975 & 3.3929101 & 105 & 1.371 & 0.466 & 52.7 & 8.0 & 0.0967 & 22.65423 & 21214 & 136.9057106 & 3.3982515 & M & 6.453 & 2.606 & 22.1 & 1 & 8.62 \\
25685 & PGC025685 & 137.0491155 & -1.8020931 & 138 & 0.428 & 0.184 & 163.4 & 19.5 & 1.0140 & 140.58021 & 3984 & 137.0486432 & -1.8032179 & M & 38.611 & 5.637 & 4.4 & 1 & -99.00 \\
25717 & UGC04804 & 137.1563580 & -1.7532529 & 0 & 0.394 & 0.369 & 113.7 & 15.1 & 2.1789 & 22.11511 & 15567 & 137.1572132 & -1.7515276 & M & 7.480 & 2.275 & 6.9 & 0 & -99.00 \\
25724 & PGC025724 & 137.1742710 & 0.4847013 & 0 & 0.347 & 0.340 & 87.1 & 18.5 & 0.3183 & 57.33348 & 30397 & 137.1741986 & 0.4825620 & M & 6.116 & 2.692 & 7.7 & 1 & 8.65 \\
25774 & PGC025774 & 137.3208900 & -1.1663546 & 76 & 0.447 & 0.183 & 113.9 & 15.2 & 4.4745 & 63.39276 & 10968 & 137.3195559 & -1.1664483 & M & 16.138 & 4.055 & 4.8 & 3 & -99.00 \\
25789 & PGC025789 & 137.3547690 & -1.7422936 & 87 & 0.284 & 0.195 & 117.2 & 15.9 & 2.8564 & 24.70218 & 13147 & 137.3532943 & -1.7432404 & M & 6.736 & 2.088 & 6.3 & 0 & -99.00 \\
25986 & UGC04857 & 138.3057270 & 3.2304294 & 57 & 0.655 & 0.256 & 49.5 & 6.4 & 0.3492 & 2.09377 & 24454 & 138.3018759 & 3.2273082 & M & 7.810 & 3.031 & 17.8 & 0 & 8.53 \\
26258 & IC0531 & 139.4616885 & -0.2784465 & 59 & 0.397 & 0.210 & 68.6 & 10.2 & 2.5077 & 43.03371 & 27962 & 139.4631403 & -0.2777034 & M & 4.222 & 2.155 & 5.9 & 0 & 8.70 \\
26385 & PGC026385 & 139.9798860 & 0.9451266 & 172 & 0.354 & 0.231 & 87.1 & 13.2 & 0.3213 & 42.73745 & 10593 & 139.9802060 & 0.9464956 & M & 11.539 & 2.962 & 5.1 & 2 & 8.63 \\
135718 & PGC135718 & 139.9799205 & -0.5913250 & 55 & 0.234 & 0.196 & 120.7 & 17.2 & 0.4388 & 2.60616 & 31226 & 139.9807500 & -0.5959474 & S & 3.619 & 2.071 & 16.9 & 0 & 8.43 \\
26398 & UGC04956 & 140.0090385 & 1.0382932 & 14 & 0.993 & 0.566 & 85.4 & 14.0 & 0.1521 & 49.97199 & 150 & 140.0090274 & 1.0388675 & M & 337.143 & 16.696 & 2.1 & 1 & 8.75 \\
26517 & UGC04978 & 140.5258800 & 3.8972574 & 0 & 0.570 & 0.541 & 57.5 & 8.6 & 0.2568 & 1.48468 & 12847 & 140.5221193 & 3.8924516 & M & 8.957 & 3.110 & 21.9 & 5 & 8.82 \\
3355297 & 2MASXJ09223417+0305019 & 140.6422605 & 3.0836643 & 62 & 0.227 & 0.205 & 50.2 & 7.7 & 0.1780 & 2.73558 & 26737 & 140.6408884 & 3.0829348 & M & 4.937 & 2.261 & 5.6 & 0 & 8.44 \\
26556 & NGC2858 & 140.7291780 & 3.1569282 & 116 & 0.879 & 0.420 & 50.5 & 7.7 & 0.2821 & 56.97719 & 29297 & 140.7284845 & 3.1570105 & M & 5.416 & 2.178 & 2.5 & 1 & 8.64 \\
26607 & NGC2861 & 140.9020770 & 2.1364922 & 0 & 0.638 & 0.583 & 69.4 & 10.0 & 2.3898 & 79.89760 & 8744 & 140.9010135 & 2.1359889 & M & 15.608 & 3.914 & 4.2 & 2 & 8.74 \\
26617 & PGC026617 & 140.9289165 & 1.3492513 & 34 & 0.454 & 0.314 & 106.8 & 13.5 & 2.6751 & 121.77942 & 24020 & 140.9298975 & 1.3465768 & M & 6.137 & 2.508 & 10.3 & 2 & 8.70 \\
135721 & PGC135721 & 140.9431110 & 2.7529378 & 160 & 0.306 & 0.131 & 69.3 & 9.9 & 0.1019 & 0.65013 & 2671 & 140.9471979 & 2.7490836 & M & 3.457 & 2.319 & 20.2 & 0 & 8.39 \\
26622 & PGC026622 & 140.9500950 & 2.1127353 & 0 & 0.378 & 0.291 & 103.1 & 12.9 & 2.3347 & 27.44695 & 22198 & 140.9522788 & 2.1081030 & M & 11.652 & 3.470 & 18.4 & 0 & 8.76 \\
1136566 & PGC1136566 & 141.3925440 & -0.7139496 & 34 & 0.112 & 0.085 & 105.6 & 13.0 & 0.1448 & 0.54200 & 11539 & 141.3937917 & -0.7160841 & M & 12.649 & 3.513 & 8.9 & 0 & 8.34 \\
26738 & NGC2877 & 141.4456965 & 2.2290502 & 0 & 0.333 & 0.320 & 103.9 & 12.9 & 9.3450 & 75.46449 & 22226 & 141.4460012 & 2.2287436 & M & 4.584 & 2.567 & 1.6 & 0 & 8.89 \\
26739 & NGC2878 & 141.4477335 & 2.0896330 & 173 & 0.518 & 0.184 & 104.2 & 13.0 & 1.0230 & 132.59200 & 12308 & 141.4489630 & 2.0875089 & M & 14.146 & 3.882 & 8.8 & 3 & 8.76 \\
26762 & UGC05027 & 141.5719080 & 3.1348488 & 172 & 0.460 & 0.091 & 67.0 & 6.9 & 0.6218 & 5.26561 & 15208 & 141.5719005 & 3.1352167 & M & 11.148 & 3.524 & 1.3 & 0 & 8.52 \\
26950 & NGC2898 & 142.4430270 & 2.0644572 & 123 & 0.518 & 0.336 & 105.7 & 32.1 & 1.8314 & 197.63673 & 475 & 142.4431924 & 2.0643592 & M & 120.669 & 9.955 & 0.7 & 2 & 8.65 \\
26988 & PGC026988 & 142.6066110 & -1.8565268 & 72 & 0.443 & 0.106 & 101.0 & 12.7 & 0.3852 & -99.00000 & 29429 & 142.6051572 & -1.8551572 & M & 6.640 & 2.430 & 7.2 & 0 & -99.00 \\
1212621 & PGC1212621 & 142.6546095 & 1.9956796 & 85 & 0.223 & 0.197 & 171.4 & 36.3 & 0.8007 & 113.16700 & 13086 & 142.6546733 & 1.9965081 & M & 14.670 & 3.933 & 3.0 & 5 & 8.76 \\
27049 & UGC05075 & 142.9234440 & 4.5000376 & 115 & 0.417 & 0.338 & 63.7 & 13.5 & 0.4600 & 65.18496 & 17593 & 142.9230803 & 4.5003034 & M & 9.879 & 3.205 & 1.6 & 1 & 8.67 \\
5953160 & SDSSJ093410.62+001431.8 & 143.5442805 & 0.2421852 & 45 & 0.439 & 0.290 & 68.2 & 12.6 & 6.0230 & 26.66908 & 3139 & 143.5449242 & 0.2414785 & M & 38.661 & 5.937 & 3.4 & 0 & 8.37 \\
27297 & PGC027297 & 143.9687790 & 4.0167868 & 151 & 0.296 & 0.217 & 75.0 & 10.6 & 0.5128 & 3.87649 & 18895 & 143.9672046 & 4.0163758 & M & 8.244 & 2.944 & 5.8 & 0 & 8.77 \\
1257748 & PGC1257748 & 144.1441680 & 3.7412125 & 5 & 0.181 & 0.136 & 204.2 & 25.3 & 1.0235 & 171.61809 & 11054 & 144.1437930 & 3.7412903 & M & 14.051 & 3.761 & 1.4 & 3 & 8.77 \\
27344 & PGC027344 & 144.1473570 & 1.1166083 & 110 & 0.358 & 0.233 & 67.8 & 9.4 & 1.4133 & 13.25034 & 20952 & 144.1508227 & 1.1130830 & M & 6.128 & 2.874 & 17.8 & 0 & 8.75 \\
27422 & NGC2936 & 144.4337400 & 2.7605731 & 77 & 0.472 & 0.259 & 102.1 & 12.8 & 13.2703 & 203.55700 & 3372 & 144.4340571 & 2.7600455 & M & 58.381 & 8.816 & 2.2 & 0 & 8.67 \\
27423 & NGC2937 & 144.4376055 & 2.7473848 & 22 & 0.376 & 0.271 & 97.8 & 13.4 & 0.6257 & 134.74999 & 3371 & 144.4360358 & 2.7463456 & M & 23.242 & 5.238 & 6.8 & 5 & 8.67 \\
27477 & UGC05138 & 144.6279135 & 2.5733679 & 171 & 0.612 & 0.104 & 101.3 & 12.7 & 1.3894 & 59.40611 & 15290 & 144.6271208 & 2.5850110 & M & 11.361 & 4.006 & 42.0 & 1 & 8.42 \\
27619 & NGC2960 & 145.1515800 & 3.5770213 & 40 & 0.553 & 0.431 & 76.2 & 14.0 & 2.6791 & 129.43493 & 6582 & 145.1513709 & 3.5764521 & M & 38.679 & 8.909 & 2.2 & -99 & -99.00 \\
27622 & IC0549 & 145.1800890 & 3.9592469 & 179 & 0.300 & 0.183 & 22.7 & 4.9 & 0.0481 & 0.84491 & 7794 & 145.1794807 & 3.9583302 & M & 25.681 & 6.472 & 4.0 & 0 & 8.63 \\
\end{longtable}
\tablefoot{
Columns (1)--(11) represent galaxy characteristics from HECATE. Columns (12)--(17) represent eFEDS X-ray source characteristics from Brunner et al. (2021, submitted). Parameter values for some sources lack data and are represented by -99. Column (5): Position angle of the galaxy. Columns (6) and (7): Semi-major and semi-minor axes of the \dtf\ ellipse, respectively. Columns (8) and (9): Distance and associated uncertainty to the galaxy. Column (15): eFEDS X-ray source catalogue provenance described in Sect. \ref{sec:samp}, where M=main and S=supplementary. Column (16) and (17): eFEDS 0.5--2 keV flux and uncertainty, respectively. Column (18): Separation between HECATE galaxy and eFEDS X-ray source. Column (19): Classification flag based on availability of optical spectroscopy (see Sect. \ref{sec:mclass}), where 0=star-forming, 1=Seyfert, 2=LINER, 3=composite, 5=spectrum exists, undefined classification, 6=early-type galaxy, $-99$=no spectrum. Column (20): Metallicity derived from optical spectra, described in Sect. \ref{sec:metal}. Galaxies with no optical emission line measurements or low S/N optical emission lines are indicated by $-99.00$. \\
}
\end{landscape}
}	% end longtable

}	% tab col sep

}	% fontsize

\bibliographystyle{aa}

\end{document}